\documentclass[aps,prl,longbibliography,showpacs,twocolumn,superscriptaddress]{revtex4-2}
\usepackage{amsmath,amssymb,amsfonts,bm}
\usepackage{booktabs}
\usepackage{braket}
\usepackage{graphicx}
\usepackage{epstopdf}
\usepackage{dcolumn}
\usepackage{mathrsfs}
\usepackage{bbold}
\usepackage{dsfont}
\usepackage{float}
\usepackage[colorlinks=true,linkcolor=blue,citecolor=blue, urlcolor=blue,bookmarks=false]{hyperref}

\usepackage{tgtermes}
\usepackage{multirow}
\usepackage{ulem}

\usepackage{mathtools}
\usepackage[dvipsnames]{xcolor}
\begin{document}

\title{Engineering Helical Superconductors with Multiple Majorana Kramers Pairs via Higher-Order Rashba Spin-Orbit Coupling}
\author{Qi-Sheng Xu}
\thanks{These authors contributed equally to this work.}
\affiliation{Department of Physics and Chongqing Key Laboratory for Strongly Coupled Physics, Chongqing University, Chongqing 400044,  China}
\author{Zi-Ming Wang}
\thanks{These authors contributed equally to this work.}
\affiliation{Department of Physics and Chongqing Key Laboratory for Strongly Coupled Physics, Chongqing University, Chongqing 400044,  China}
\author{Chui-Zhen Chen}
\affiliation{Institute for Advanced Study and School of Physical Science and Technology, Soochow University, Suzhou 215006, China}
\author{Lun-Hui Hu}
\email{hu.lunhui.zju@gmail.com}
\affiliation{Center for Correlated Matter and School of Physics, Zhejiang University, Hangzhou 310058, China}

\author{Rui Wang}
\affiliation{Department of Physics and Chongqing Key Laboratory for Strongly Coupled Physics, Chongqing University, Chongqing 400044, China}
\affiliation{Center of Quantum Materials and Devices, Chongqing University, Chongqing 400044, China}

\author{Dong-Hui Xu}
\email{donghuixu@cqu.edu.cn}
\affiliation{Department of Physics and Chongqing Key Laboratory for Strongly Coupled Physics, Chongqing University, Chongqing 400044, China}
\affiliation{Center of Quantum Materials and Devices, Chongqing University, Chongqing 400044, China}
	
\begin{abstract}
The momentum dependence of Rashba spin-orbit coupling (RSOC) is a key ingredient for engineering topological superconductors (TSCs), yet research has overwhelmingly focused on its linear-in-momentum form. This focus has restricted time-reversal invariant TSCs to helical $p$-wave states, which are characterized by a $\mathbb{Z}_2$ topological invariant that permits at most a single Majorana Kramers pair at a given boundary. Their existence has also been tied to the stringent criterion of an odd number of Fermi surfaces (FSs). In this work, we establish higher-order RSOC as a powerful design principle to go beyond the $\mathbb{Z}_2$ classification and the odd-FS criterion. We demonstrate that a bilayer system with a pure cubic RSOC and an intrinsic odd-parity pairing on a single FS yields a rare 2D helical $f$-wave TSC. This state is characterized by a large mirror Chern number (MCN) of ${\cal N}_{\text{M}}=3$ and hosts three Kramers pairs of Majorana edge modes. Remarkably, the interplay of linear and cubic RSOCs in this bilayer can generate a helical hybrid $p+f$-wave TSC with an even larger MCN of ${\cal N}_{\text{M}}=4$ from a normal state with two FSs, thereby circumventing the conventional odd-FS criterion. Our work establishes higher-order RSOC as a ``topology multiplier" for realizing TSCs with  multiple Majorana Kramers channels, fundamentally reshapes the criteria for helical TSCs, and holds immediate relevance for tunable platforms like oxide heterostructures.

\end{abstract}
	
\maketitle

Spin-orbit coupling is the relativistic interaction between an electron’s spin and its orbital motion. In crystals lacking inversion symmetry, this interaction manifests as Rashba spin-orbit coupling (RSOC), which lifts the spin degeneracy of electronic bands through momentum-dependent spin splitting~\cite{bychkov1984properties}. RSOC is a cornerstone for realizing exotic quantum phenomena~\cite{manchon2015new,bihlmayer2022rashba,spinhall,TIRMP2010,TIRMP2011}. In superconductors, RSOC facilitates the admixture of spin-singlet and triplet Cooper pairs~\cite{mixsinglettriplet}, significantly enhancing the upper critical field~\cite{mixsinglettriplet,Yip02,Sigrist04}. Crucially, the influence of RSOC is not limited to globally noncentrosymmetric materials; a hidden RSOC can emerge even in centrosymmetric structures, provided that local atomic environments break inversion symmetry~\cite{zhang2014hidden,gotlieb2018revealing,khim2021field}. This phenomenon, observed in systems such as bismuth-based cuprates~\cite{gotlieb2018revealing} and heavy fermion superconductors~\cite{khim2021field}, has significantly broadened the materials landscape for novel superconducting phases, particularly those characterized by unconventional odd-parity Cooper pairing~\cite{fischer2023superconductivity}.

A primary motivation for studying RSOC is its central role in engineering topological superconductors (TSCs). While much focus has been placed on localized Majorana zero modes for braiding-based quantum computation~\cite{nayakrmp}, TSCs also host propagating Majorana edge channels. These itinerant modes are essential for ``flying" Majorana qubit architectures and interferometry-based topological quantum computation~\cite{Beenakkerscipost}. Seminal proposals have leveraged the spin-momentum locking of conventional, linear-in-momentum~($\bm{k}$) RSOC to generate TSCs. For instance, an $s$-wave superconductor proximitized to the surface of a topological insulator can realize an effective $p_x+ip_y$ state supporting Majorana zero modes in vortex cores~\cite{Fukanemodel}. By breaking time-reversal symmetry (TRS) with a magnetic field, two-dimensional (2D) systems with linear RSOC can host chiral $p$-wave TSCs~\cite{Fujimoto08,sato09,Jsau10,Alicea10}. Furthermore, linear RSOC is central to proposals for 2D helical TSCs that preserve TRS. These states belong to symmetry class DIII, and their $\mathbb{Z}_2$ topological classification restricts them to hosting at most a single robust Kramers pair of Majorana edge modes~\cite{CXLiuhelial,Nakosaihelical12,Fanzhanghelical13,zhang2013topological}. The existence of these states has also been tied to a stringent odd-FS criterion, which requires an odd number of Fermi surfaces (FSs) to enclose time-reversal invariant momenta~\cite{oddparityTscFu,Sato-odd,Qi-TRISC}. A crucial question is how to transcend this binary framework to realize 2D TSCs with multiple Kramers pairs of Majorana channels, effectively transitioning from a $\mathbb{Z}_2$ to a $\mathbb{Z}$ classification.

\begin{figure}[t]
\includegraphics[width=0.48\textwidth]{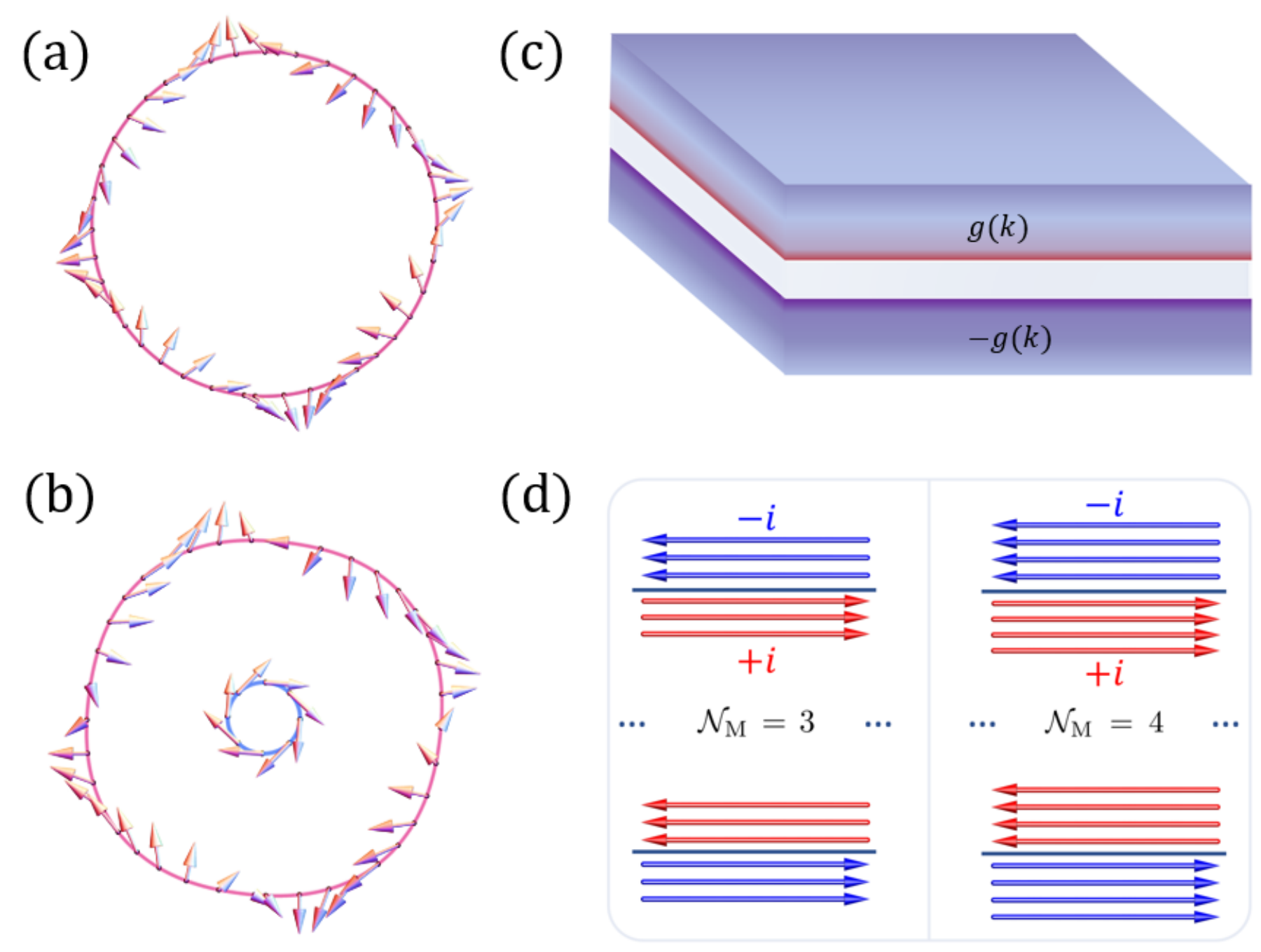 }
\caption{Spin textures and the resulting helical Majorana channels. (a) Triple spin winding on a single FS due to cubic RSOC. (b) A combination of triple (outer FS) and single (inner FS) spin windings arising from the interplay of cubic and linear RSOC. (c) Illustration of a bilayer electron gas with staggered RSOC, a physical platform realized in systems like oxide heterostructures. (d) The helical Majorana edge modes protected by mirror symmetry. The cubic RSOC case results in three pairs of channels (left), while the coexistence of linear and cubic RSOCs yields four pairs (right).}
\label{fig1}
\end{figure}

A promising avenue for overcoming this limitation involves exploring spin textures beyond the simple, single winding of linear RSOC. Recently, significant attention has shifted towards a higher-order, cubic-in-$\bm{k}$ RSOC~\cite{CRSOCoxide,CRSOCoxidedft,RSOCGe,CRSOCstotransport,CRSOCstokto,lin2019interface,CRSOCrareearth,CRSOCDFT,RSOCGe,CRSOCspinhall05,CRSOCspinhall06,CRSOCoxides,CRSOCKondo}, which can dominate when the linear term is suppressed by symmetry. This cubic RSOC is particularly relevant in oxide interfaces such as LaAlO$_3$/SrTiO$_3$~(LAO/STO), where its strength is electrically tunable~\cite{lin2019interface,caviglia2010tunable,shalom2010prl} and superconductivity is well established~\cite{reyren2007superconducting,caviglia2008electric}. The spin texture generated by cubic RSOC is markedly different from its linear counterpart. As an electron's momentum traverses a FS, its spin, locked by cubic RSOC, rotates three full times, creating a triple-winding texture~[Fig.~\ref{fig1}(a)]. While prior work has explored proximity-induced, TRS-breaking TSCs~\cite{CRSOCchiralsc,CROSJosephson1,CROSJosephson2}, the potential for a TRS-invariant TSC driven by an intrinsic pairing instability in a cubic RSOC system--and the consequent realization of multiple, controllable Majorana Kramers pairs arising from its unique triple-winding spin texture--remain key open questions. 

\begin{table*}[!htbp]
    \caption{\label{tab:table1}%
    Classification of possible momentum-independent pairing potentials from the short-range interaction model. The pairings $\Delta_1,\Delta_2,\Delta_3,$ and $\Delta_4$ belong to the $A_{1g}, A_{1u}, A_{2u}, $ and $E_u$ irreps of $D_{4h}$, respectively. Matrix representations  are given in the layer-spin basis, and transformation properties under inversion (${\cal I}$) and mirror (${\cal M}_z$) symmetries are listed.}
    \begin{ruledtabular}
    \tabcolsep=0.1cm
    \renewcommand{\arraystretch}{1}
    \begin{tabular}{ccccc}
    Pairing & Operator representation & Matrix & ${\cal I}$& ${\cal M}_z$ \\
    \hline
    $\Delta_1$ & $c_{l,\uparrow}c_{l,\downarrow}+c_{u,\uparrow}c_{u,\downarrow},\;\;c_{l,\uparrow}c_{u,\downarrow}-c_{l,\downarrow}c_{u,\uparrow}$ & {$i\tau_0 s_y$,\;\;$i\tau_x s_y$} & $+$ &$+$\\
    $\Delta_2$ & $i${$( c_{l,\uparrow}c_{u,\downarrow}+c_{l,\downarrow}c_{u,\uparrow}$)}& {$\tau_y s_x $} & {$-$} & $-$ \\
    $\Delta_3$ & {$c_{l,\uparrow}c_{l,\downarrow}-c_{u,\uparrow}c_{u,\downarrow}$} & {$i\tau_z s_y$} & {$-$} & $-$ \\
    $\Delta_4$ & {$(i(c_{l,\uparrow}c_{u,\uparrow}+c_{l,\downarrow}c_{u,\downarrow}),c_{l,\uparrow}c_{u,\uparrow}-c_{l,\downarrow}c_{u,\downarrow})$}  & {($\tau_y s_z, i\tau_y s_0$)} & {$-$} & $+$ 
    \end{tabular}
    \end{ruledtabular}
\end{table*} 

In this work, we establish a direct mechanism for realizing novel TRS invariant helical TSCs that host multiple Kramers pairs of Majorana channels. We study an interacting bilayer model with local inversion symmetry breaking--a canonical platform for hidden RSOC. We find that the triple-winding spin texture inherent to cubic RSOC, combined with an odd-parity pairing instability on a single FS, leads to a TSC protected by mirror symmetry. The bulk topology is characterized by a large mirror Chern number~(MCN), ${\cal N}_{\text{M}}\!=\!3$, guaranteeing three Kramers pairs of Majorana edge modes~[Fig.~\ref{fig1}(d), left panel]. Furthermore, we show that the coexistence of linear and cubic RSOCs can generate a helical hybrid $p+f$-wave TSC characterized by ${\cal N}_{\text{M}}\!=\!4$, hosting four Kramers pairs of Majorana edge modes~[Fig.~\ref{fig1}(d), right panel]. Crucially, this state emerges from a normal state with two FSs~[Fig.~\ref{fig1}(b)], thereby circumventing the established odd-FS criterion for helical TSCs. Our results establish a direct link between the winding of the normal-state spin texture and the number of protected Majorana channels, opening a new route towards engineering topological phases with tailored properties.

\emph{Model.}---We consider a minimal model for locally noncentrosymmetric structures, realized by a 2D bilayer system possessing $D_{4h}$ point group symmetry. While the system as a whole possesses global inversion symmetry, this symmetry is broken locally within the individual layers. This local asymmetry induces a hidden spin polarization~\cite{zhang2014hidden,gotlieb2018revealing,khim2021field}, which manifests as a staggered RSOC in the bilayer geometry, as depicted in Fig.~\ref{fig1}(c). The low-energy normal-state Hamiltonian describing this system is
\begin{align}
{\cal H}_{\text{N}}(\bm{k})
&= \epsilon_0(\bm{k}) \tau_0 \otimes s_0 + \tau_z \otimes [ \bm{g}(\bm{k}) \cdot \bm{s} ] + \varepsilon \tau_x \otimes s_0,    
\label{eq1}
\end{align} 
where the Pauli matrices $\tau_{x,y,z}$ and $s_{x,y,z}$ act on the layer {lower ($l$), upper ($u$)} and spin {$\uparrow$, $\downarrow$} degrees of freedom, respectively, and $\tau_0$ and $s_0$ are the corresponding $2\times2$ identity matrices. The first term, $\epsilon_0(\bm{k})\!\!=\!\!\beta_0 {k}^4\!+\!\beta_1 {k}^2$, describes the intralayer kinetic energy, where the quartic term can account for nonparabolic band effects. The second term represents the staggered RSOC, where $\tau_z$ ensures the RSOC has opposite signs in the two layers. The RSOC vector $\bm{g}(\bm{k})\!\!=\!\!\bm{g}_{\text{LR}}(\bm{k})\!+\!\bm{g}_{\text{CR}}(\bm{k})$ includes terms consistent with $D_{4h}$ symmetry, which are linear and cubic in momentum $\bm{k}$: $\bm{g}_{\text{LR}}(\bm{k})\!\!=\!\!\alpha_\text{LR}(-{k}_y,{k}_x,0)$ and $\bm{g}_{\text{CR}}(\bm{k})\!\!=\!\!\alpha_\text{CR} [-{k}_y(3{k}_x^2-{k}_y^2),{k}_x(3{k}_y^2-{k}_x^2),0]$. The third term, with strength $\varepsilon$, is the momentum-independent interlayer tunneling, which opens a hybridization gap~(HG) of $2\varepsilon$  around the $\Gamma$ point~[Fig.~\ref{fig2}(a)]. The normal Hamiltonian, ${\cal H}_{\text{N}}(\bm{k})$, preserves several key symmetries: TRS represented by the anti-unitary operator $\mathcal{T}\!\!=\!\!-i s_y {\cal K}$ (where ${\cal K}$ is the complex conjugate), inversion symmetry ${\cal I}\!\!=\!\!\tau_x$, fourfold rotational symmetry $\quad {\cal R}_{4z} \!\!=\!\! i \tau_x e^{i \tfrac{\pi}{2} \tfrac{s_z}{2}}$  and  mirror symmetry $\mathcal{M}_z\!\!=\!\!i\tau_xs_z$.

\begin{figure}[!htbp]
\centering
\includegraphics[width=0.48\textwidth]{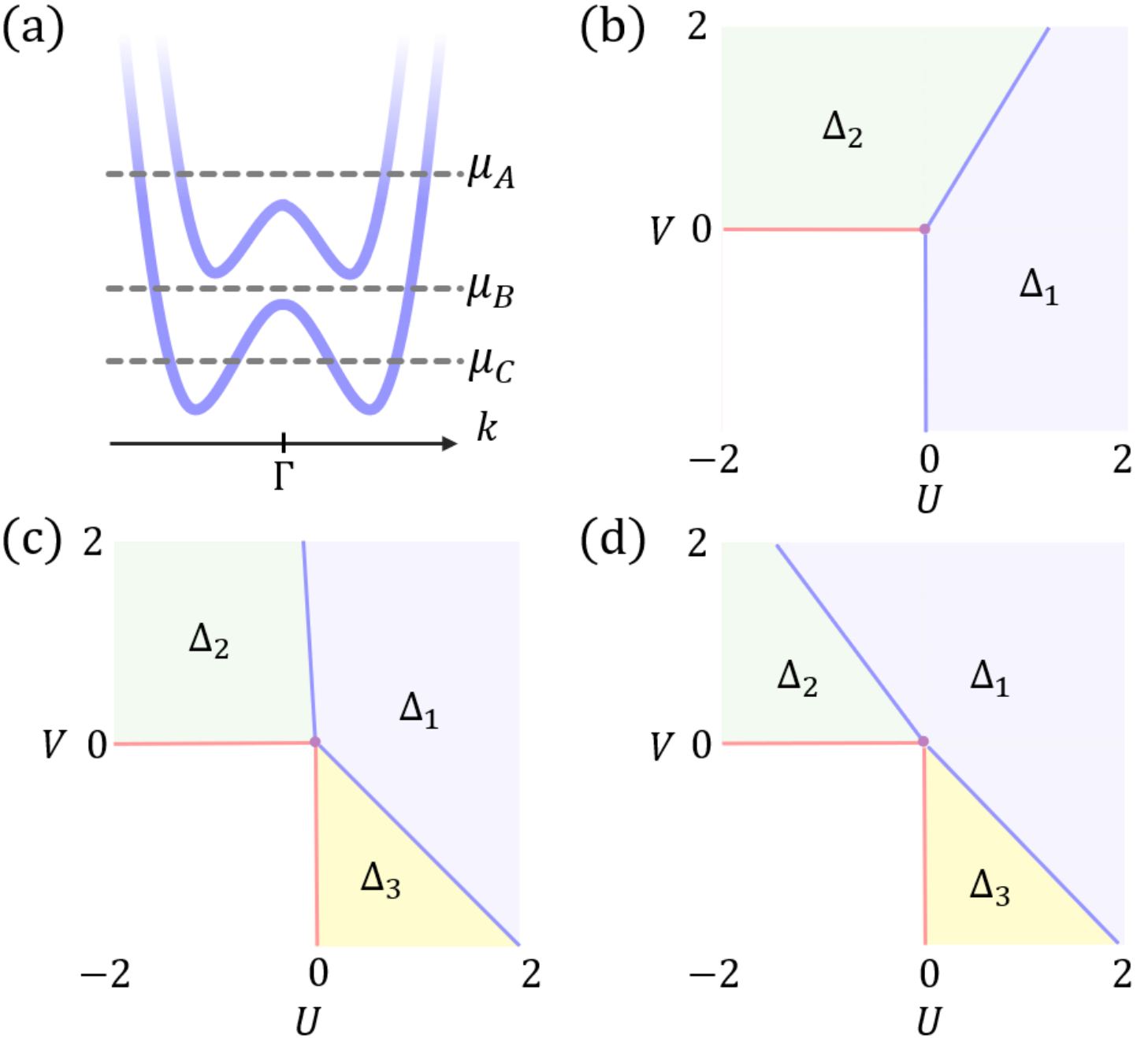}
\caption{ (a) Energy dispersion at the $\Gamma$, illustrating the HG induced by 
$\varepsilon$. The chemical potential $\mu_A=1.05$ intersects both upper and lower bands, while $\mu_B=-0.25$ and $\mu_C=-0.4$ intersect only the lower band. (b-d) Superconducting phase diagram for each chemical potential. The phases are color-coded as follows: $\Delta_1$ (pale blue), $\Delta_2$ (pale green), $\Delta_3$ (pale yellow), and a non-superconducting phase (white). Common parameters, $\beta_0\!\!=\!\!0.5$, $\beta_1\!\!=\!\!-1$, $\varepsilon\!\!=\!\!0.35$, $\alpha_\text{CR}\!\!=\!\!0.1$.
}
\label{fig2}
\end{figure}

To investigate superconductivity, we introduce a phenomenological short-range density-density interaction, which has proven effective for studying competing pairing instabilities in various multi-orbital and multilayer systems~\cite{oddparityTscFu,Nakosaihelical12,Sato-DSM}. The interaction Hamiltonian in real space is
\begin{align} \label{eq-ham-int}
{\cal H}_\text{Int}(\bm{r}) = -U[n_l^2(\bm{r})+n_u^2(\bm{r})]-2Vn_l(\bm{r})n_u(\bm{r}),
\end{align}
where $n_{\tau}(\bm{r})\!\!=\!\!\sum_{s}c^\dagger_{\tau,s}(\bm{r})c_{\tau,s}(\bm{r})$ is the local electron density operator in the lower ($\tau\!\!=\!\!l$) and upper ($\tau\!\!=\!\!u$) layers. The parameters $U$ and $V$ represent the effective strengths of the intra-layer and inter-layer interactions, respectively; positive values correspond to attractive interactions. While introduced phenomenologically, their values are grounded in the physics of dilute oxide superconductors. In STO-based systems, pairing is mediated by soft phonons competing with screened Coulomb repulsion~\cite{gor2016phonon}. The strong dielectric screening in the plane typically results in a net attractive intralayer potential ($U>0$), while the less effective screening across the barrier leaves the interlayer interaction repulsive or weakly renormalized ($V<0$). These effective parameters are experimentally tunable via gate voltage, see Supplemental Material (SM)~\cite{Supp}.

 We then solve $\cal H_{\text{N}}\!\!+\!\! \cal H_{\text{Int}}$ using the standard mean-field approximation. The interaction term is decoupled in the particle-particle (Cooper) channel, leading to momentum-independent pairing potentials whose symmetries are constrained by $D_{4h}$. To formalize this, we introduce an eight-component Nambu spinor, defined as $\psi_{\bm{k}}^\dagger\!\!=\!\!(c_{l\bm{k},\uparrow}^\dagger,c_{l\bm{k},\downarrow}^\dagger,c_{u\bm{k},\uparrow}^\dagger,c_{u\bm{k},\downarrow}^\dagger,c_{l-\bm{k},\uparrow}, c_{l-\bm{k},\downarrow},c_{u-\bm{k},\uparrow},c_{u-\bm{k},\downarrow})$. The Bogoliubov-de Gennes (BdG) Hamiltonian takes the block form 
\begin{align}
{\cal H}_\text{BdG} ({\bm k}) =\begin{pmatrix}
 {\cal H}_{\text{N}}(\bm{k})-\mu & \Delta(\bm{k}) \\
 \Delta^\dagger(\bm{k}) & -{\cal H}^\ast_{\text{N}}(-\bm{k})+\mu
\end{pmatrix},
\label{sseq3}
\end{align}
where $\mu$ is the chemical potential, and $\Delta(\bm{k})$ is the $4\times 4$ pairing potential matrix in the layer-spin space. The dominant pairing symmetries derived from the intralayer ($U$) and interlayer ($V$) interactions are summarized in Table~\ref{tab:table1}. The even-parity channel, transforming as the $A_{1g}$ irrep, is a spin-singlet state. It contains two components: an intralayer pairing $\Delta_{1a}\!\!=\!\!\Delta_0 ( c_{l,\uparrow}c_{l,\downarrow}+c_{u,\uparrow}c_{u,\downarrow} )$ and an interlayer pairing $\Delta_{1b}\!\!=\!\!\Delta_0 (c_{l,\uparrow}c_{u,\downarrow}-c_{l,\downarrow}c_{u,\uparrow})$. Here, $\Delta_0$ is the strength of the pairing potential. Because they share the same symmetry, these two components can mix. The remaining channels are all odd-parity and correspond to spin-triplet states. The $\Delta_2$~($A_{1u}$) and $\Delta_3$~($A_{2u}$) channels represent interlayer and intralayer pairings, respectively. $\Delta_{4x}\!\!=\!\!i\Delta_0 (c_{l,\uparrow}c_{u,\uparrow}+c_{l,\downarrow}c_{u,\downarrow})$ and $\Delta_{4y}\!\!=\!\!\Delta_0 (c_{l,\uparrow}c_{u,\uparrow}-c_{l,\downarrow}c_{u,\downarrow})$  belong to the 2D $E_u$ irrep and describes an interlayer equal-spin pairing state.

\emph{Linearized gap equations and phase diagram.}---For the subsequent analysis, we focus on the physics driven by the higher-order RSOC term. Therefore, we initially simplify the model by setting the linear Rashba coefficient to zero ($\alpha_\text{LR}\!\!=\!\!0$), isolating the effects of the cubic RSOC. In the weak-coupling limit, we can determine the superconducting critical temperature $T_c$ by solving the linearized gap equation, which can be expressed using the pairing susceptibility. The pairing susceptibility $\chi$ for a given pairing symmetry is calculated from the particle-particle correlation function
\begin{equation}
    \begin{aligned}\chi_{i}
           &\equiv-\frac{1}{\beta_c}\sum_{\omega_n,\bm{k}}\text{Tr}\Big[\Gamma^\dagger_{i}\mathcal{G}_e(\omega_n,\bm{k})\Gamma_{i}\mathcal{G}_h(\omega_n,\bm{k})\Big]\\
           &=-\sum_{\bm{k}, \eta=\pm}\Bigg [ F_{i}^{\eta}(\bm{k})\frac{\tanh[\beta_c(E_\eta-\mu)/2]}{2(E_\eta-\mu)}\Bigg ],
    \label{seq4} 
    \end{aligned}
\end{equation}
where $\omega_n$ are the fermionic Matsubara frequencies, $\beta_c=1/(k_B T_c)$, and $\mathcal{G}_{e,h}$ are the normal Green's functions for electrons and holes, respectively. The matrices $\Gamma_{i}$ are the vertex functions corresponding to the pairing channel whose forms are given in Table~\ref{tab:table1}. After performing the Matsubara summation, the susceptibility is expressed as a sum over the normal state energy bands, $E_{\pm}(\bm{k})\!\!=\!\!\epsilon_0 (\bm{k})\pm \lambda_{\bm{k}}$, where $\lambda_{\bm{k}}\!\!=\!\!\sqrt{\varepsilon^2+\alpha_\text{CR}^2 k^6}$. The coefficients $F^{\pm}_{ij}$ depend on the projection of the pairing vertices onto the eigenvectors of the normal state Hamiltonian and are detailed in the SM~\cite{Supp}.

The linearized gap equations for the different pairing channels are then constructed. For the even-parity $A_{1g}$ channel ($\Delta_1$), the intralayer ($\Delta_{1a}$) and interlayer ($\Delta_{1b}$) components belong to the same irrep and are thus coupled, leading to a matrix equation
\begin{align}
\begin{vmatrix}
 U\chi _{1a,1a}-1 & U \chi _{1a,1b} \\
 V\chi _{1b,1a} & V\chi _{1b,1b}-1
\end{vmatrix}=0.
\label{seq5}
\end{align}
For the odd-parity channels $\Delta_2$~($A_{1u}$), $\Delta_3$~($A_{2u}$), and $\Delta_4$~($E_u$), which belong to distinct irreps, the gap equations are decoupled
\begin{align}
 V \chi_2=1, U \chi_3=1, V \chi_4=1.
 \label{seq6}
\end{align}

By solving these equations, the superconducting phase diagram in the $U$-$V$ plane is determined by identifying the pairing channel with the maximum critical temperature $T_c$. Figures~\ref{fig2}(b)-\ref{fig2}(d) show the resulting phase diagrams for three representative $\mu$: $\mu_A$, which intersects both the upper and lower bands, while $\mu_B$ and $\mu_C$ intersect only the lower band, with $\mu_B$ located within the HG~[Fig.~\ref{fig2}a]. The phase diagram reveals a rich landscape where three distinct superconducting states--$\Delta_1$, $\Delta_2$, and $\Delta_3$--compete for dominance against normal metallic phase. $\Delta_4$ never emerges as the leading instability for the interactions considered.

The outcome of this competition is dictated by the interplay between the intralayer ($U$) and interlayer ($V$) interactions. Specifically, a dominant attractive intralayer interaction ($U>0$) stabilizes the conventional spin-singlet state $\Delta_1$. Conversely, when the interlayer interaction prevails $(|V|>U)$, the system favors one of two unconventional spin-triplet phases, $\Delta_2$ or $\Delta_3$. $\Delta_2$ phase can be stabilized by a combination of repulsive intralayer and attractive interlayer interactions; however, this stabilization requires dominant $V$ for $\mu_C$. If both interactions are repulsive, no superconducting instability occurs, and the system remains in a normal metal.
The intralayer odd-parity $\Delta_3$ is of primary focus of this work. Its stability region appears in cases $\mu_B$ and $\mu_C$. As seen in Figs.~\ref{fig2}(c) and \ref{fig2}(d), the parameter regime with $U>0$ and $V<0$ strongly favors the $\Delta_3$ pairing state, which is experimentally relevant and potentially accessible in oxide-interface platforms.
%The emergence of this odd-parity state as a ground state is plausible in realistic material settings, such as oxide interfaces, where superconductivity arises from a competition between screened Coulomb repulsion and phonon-mediated attraction. The parameters $U$ and $V$ represent the net effective interactions resulting from this interplay({\color{blue} See SM for more discussion on interactions}). It's reasonable to assume that both screening and phonon coupling are more effective within the layers than between them. This asymmetry can create a scenario where the effective intralayer interaction $U$ becomes attractive ($U>0$) while the effective interlayer interaction $V$ remains repulsive ($V<0$). 

In 2D, the linearized gap equation determines the pairing scale ($T^*$), while the finite-temperature transition is of BKT type with ($T_\text{BKT}<T^*$). Since our topological invariants and edge spectra are properties of the  $T=0$ BdG Hamiltonian, they are correctly captured by mean-field; see SM Sec. III for the regime-of-validity discussion~\cite{Supp}.

\emph{Helical $f$-wave TSC.}---We now analyze the topological properties of the superconducting phase driven by the order parameter $\Delta_3$. The pairing $\Delta_3$ is odd under the mirror operation ${\cal M}_z$. Accordingly, the BdG Hamiltonian remains invariant under the Nambu-extended mirror symmetry $\tilde{\cal M}_z\equiv\text{diag}[{\cal M}_z, -{\cal M}^{\ast}_z]$, which enables block diagonalization into the $\pm i$ mirror sectors. The BdG Hamiltonian for each sector is ${\cal H}_{\pm i}(\bm{k})$, where 
\begin{align*}
{\cal H}_{\pm i}(\bm{k})=\begin{pmatrix}
 {\cal H}_{\text{N},\pm i}(\bm{k})-\mu &-i\Delta_0 s_y  \\
 i \Delta_0 s_y  & -{\cal H}^\ast_{\text{N},\pm i}(-\bm{k})+\mu 
\end{pmatrix},
\end{align*}
with ${\cal H}_{\text{N},\pm i}(\bm{k})\!\!=\!\!-\epsilon_0(\bm{k}) s_0 -\bm{g}_{\text{CR}}(\bm{k}) \cdot \bm{s} \pm \varepsilon s_z$, and $\mu_B$ located within the HG. ${\cal H}_{+i}(\bm{k})$ and ${\cal H}_{-i}(\bm{k})$ lack TRS individually, but are mapped onto each other by TRS operator $\mathcal{T}$. To reveal the effective pairing symmetry induced by $\Delta_3$, we project the Hamiltonian ${\cal H}_{\pm i}$ onto the basis of the normal-state spin-split bands, ${\cal H}_{\text{N},\pm i}$. Focusing on the $+i$ block, we project the $\Delta_3$ pairing potential (which is proportional to $-is_y$ in this subspace) onto the upper ($+$) and lower ($-$) bands. The projected potentials are~\cite{Supp}
\begin{equation}
\begin{aligned}
&\Delta_{++}(\bm{k})=-\Delta_0\frac{\alpha_\text{CR}}{\lambda_{\bm{k}}}\Big(-ik_x+k_y\Big)^3,\\
&\Delta_{+-}(\bm{k})=-\Delta_0\frac{\varepsilon}{\lambda_{\bm{k}}}, \\
&\Delta_{--}(\bm{k})=-\Delta_0\frac{\alpha_\text{CR}}{\lambda_{\bm{k}}}\Big(ik_x+k_y\Big)^3.
\end{aligned}
\end{equation}
This result is central to our work: the triple-winding spin texture of the normal state, encoded in the cubic momentum dependence of $\bm{g}_{\text{CR}}(\bm{k})$, is directly inherited by the intraband pairing potential. The resulting gap functions, $\Delta_{++} (\bm{k})\propto (k_y-ik_x)^3$ and $\Delta_{--} (\bm{k}) \propto (k_y+ik_x)^3$ correspond to $f_{y(3x^2-y^2)}\mp i f_{x(x^2-3y^2)}$ symmetries with opposite chiralities for the upper and lower bands. When $\mu$ lies within the HG~[Fig.~\ref{fig2}(a)], only the lower band crosses $\mu$. The system is then effectively a single band superconductor. In this regime, the ${\cal H}_{+i}$ sector describes a chiral $f$-wave TSC. The ${\cal H}_{-i}$ sector describes the time-reversed partner with opposite chirality. Together, they form a TRS-invariant TSC with helical $f$-wave pairing protected by mirror symmetry ${\cal M}_z$.

\begin{figure}[!htbp]
\centering
\includegraphics[width=0.48\textwidth]{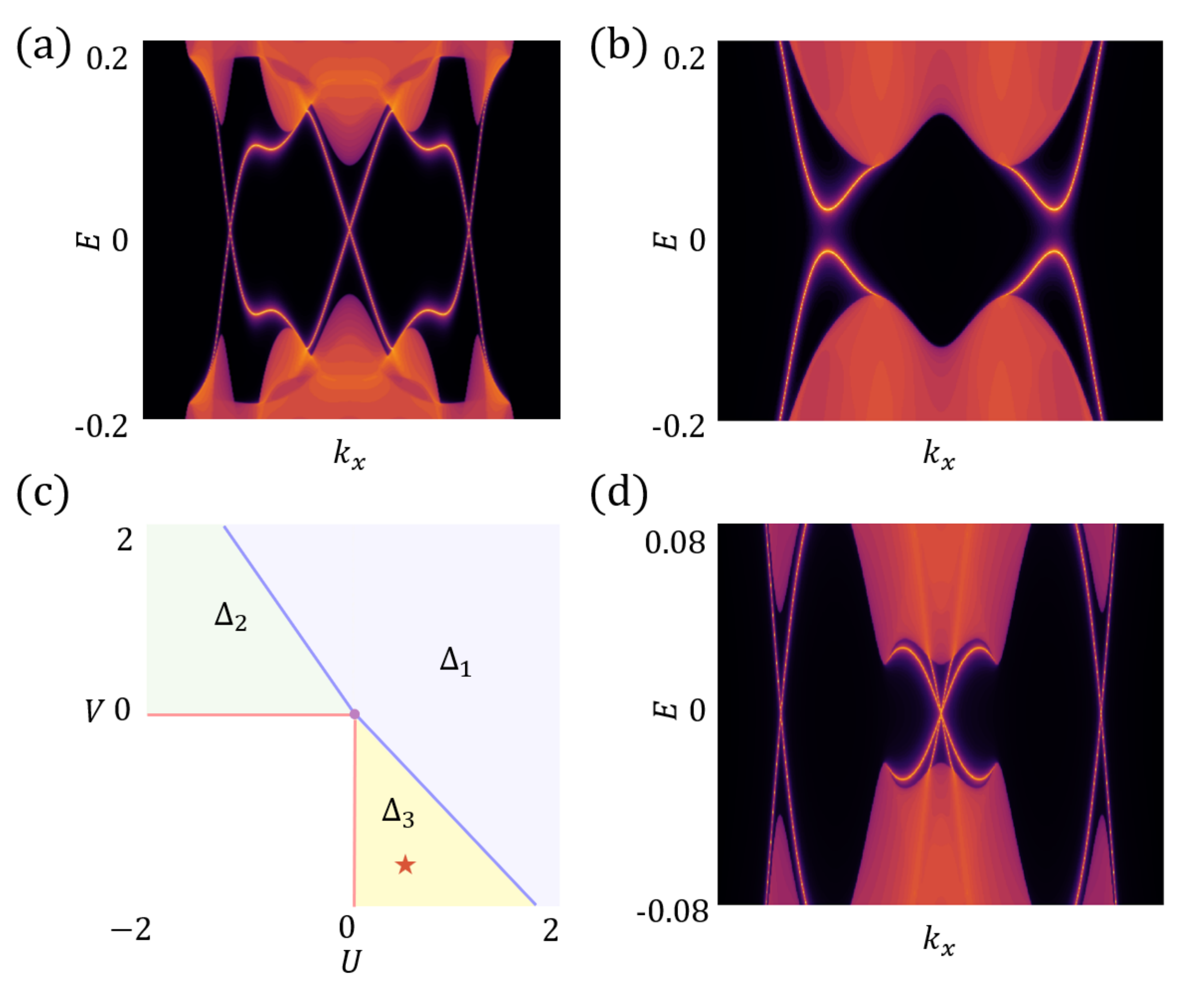}
\caption{Superconducting phase diagram and spectral function ${\cal A}(E,\bm{k})$ for the $\Delta_3$ pairing phase in a semi-infinite geometry. (a, b)  ${\cal A}(E,k_x)$ for the pure cubic RSOC case ($\alpha_\text{CR}=0.1$). Three Kramers pairs of Majorana edge modes appear when the chemical potential is in HG as in (a) $\mu_B=-0.2$, but not outside it as in (b) $\mu_C=-0.6$. (c) Phase diagram of coexisting linear and cubic RSOCs $\alpha_\text{CR}=\alpha_\text{LR}=0.1$ at $\mu_C=-0.4$. (d) ${\cal A}(E,k_x)$ at the point marked by red star in (c), revealing four pairs of helical Majorana modes. Common parameters: $\beta_0=0.5, \beta_1=-1, \varepsilon=0.35, \Delta_0=0.2$.} 
\label{fig3}
\end{figure}

 According to the topological classification of reflection-symmetric superconductors, our Rashba bilayer belongs to class DIII+$R_{--}$~\cite{chiu2013classification}. Its bulk topology is characterized by the
MCN ${\cal N}_{\text{M}}\!\!=\!\!({\cal N}_{+i}\!-\!{\cal N}_{-i})/2$, where ${\cal N}_{\pm i}$ are the Chern numbers evaluated in the $\mathcal{M}_z=\pm i$ mirror subsectors
~\cite{PhysRevB.78.045426,ueno2013symmetry,chiu2013classification,tsutsumi2013upt3,shiozaki2014topology,ando2015topological,yoshida2015topological}. A direct calculation shows that when $\mu$ is inside the HG and the system is fully gapped (requiring $\varepsilon^2>\mu^2+\Delta_0^2$), the MCN is ${\cal N}_{\text{M}}\!\!=\!\!3$. The magnitude ${\cal N}_{\text{M}}$ directly reflects the triple-winding topology of the cubic Rashba spin texture~[Fig.~\ref{fig1}(a)], which has been faithfully transferred to the superconducting order parameter. This demonstrates the ``topology multiplier" principle: the threefold winding of the spin texture generates a threefold enhancement of the topological invariant. The bulk-boundary correspondence dictates that a TSC with MCN ${\cal N}_{\text{M}}$ will host $|{\cal N}_{\text{M}}|$ pairs of counterpropagating Majorana edge modes. To verify this, we discretize the model on a square lattice and calculate the spectral function ${\cal A}(E,\bm{k})$ for a semi-infinite geometry. As shown in Fig.~\ref{fig3}(a), ${\cal A}(E,\bm{k})$ for the $\Delta_3$ phase with ${\cal N}_{\text{M}}\!\!=\!\!3$ clearly shows three pairs of linearly dispersing, gapless modes crossing zero energy. These are the predicted three helical Majorana channels, in stark contrast to the single pair expected from linear RSOC~\cite{CXLiuhelial,Nakosaihelical12,Fanzhanghelical13}. 

When the chemical potential is lowered to $\mu_C$, the system has two concentric FSs enclosing the $\Gamma$ point. In this case, even though the $\Delta_3$ pairing state is still favored for $U>0$ and $V<0$, the system is topologically trivial, as shown by the fully gapped edge spectrum in Fig.~\ref{fig3}(b). This is a direct consequence of the even number of FSs, consistent with the established criterion for odd-parity superconductors~\cite{Sato-odd,oddparityTscFu}. However, this classification is fundamentally altered when cubic and linear RSOCs coexist.

\emph{Helical hybrid $p+f$-wave TSC.}---A critical test for experimental relevance is the stability of a topological phase against perturbations expected in real materials. For instance, in oxide heterostructures like LAO/STO~\cite{lin2019interface,caviglia2010tunable,shalom2010prl,CRSOCoxide}, both linear and cubic RSOCs coexist, with their relative strengths tunable via carrier filling. Reported values are typically $\alpha_\text{LR} \!\approx\! 10-50$ meV$\cdot$\AA\; and $\alpha_\text{CR} \!\approx\! 1-10$ eV$\cdot$\AA$^3$ for the linear and cubic terms. These coefficients imply a crossover momentum $k_c= \sqrt{\alpha_\text{LR}/\alpha_\text{CR}}$ corresponding to carrier densities of order $10^{13}$ cm$^{-2}$, a regime well within the experimentally accessible window for oxide interfaces. We therefore investigated the robustness of the ${\cal N}_{\text{M}}\!\!=\!\!3$ phase by reintroducing the linear RSOC. We found that the three pairs of helical Majorana edge modes persist for dominant cubic RSOC as long as $\mu$ lies within HG (See SM~\cite{Supp}).

A remarkable new phenomenon emerges when $\mu$ is moved out of HG but remains within the lower band. This creates two distinct FSs: a smaller, inner surface near the $\Gamma$ point and a larger, outer surface farther away~[Fig.~\ref{fig1}(b)]. In this regime $\Delta_3$ pairing remains favored under similar interaction conditions~[Fig.~\ref{fig3}(c)]. The interplay between coexisting linear and cubic RSOCs and $\Delta_3$ pairing induces an even richer TSC characterized by a heightened MCN, ${\cal N}_{\text{M}}\!\!=\!\!4$, hosting four Kramers pairs of Majorana edge modes~[Fig.~\ref{fig3}(d)]. This state is a beautiful example of $\bm{k}$-space topological engineering. The different momentum scaling of the two RSOCs allows them to dominate in different regions of the Brillouin zone. Given sufficient separation of the FSs in ${\bm k}$-space, the linear term ($\propto k$) dominates the inner surface, fostering a helical $p$-wave pairing that contributes ${\cal N}_{\text{inner}}\!\!=\!\!1$ to the MCN. Conversely, the cubic term ($\propto k^3$) dominates the outer surface, fostering a helical $f$-wave pairing that contributes ${\cal N}_{\text{outer}}\!\!=\!\!3$. Since the MCN is an additive integer invariant, the total MCN for the system is ${\cal N}_{\text{M}}\!\!=\!\!{\cal N}_{\text{inner}}+{\cal N}_{\text{outer}}=1+3=4$. This result is profound: it demonstrates the realization of a robust TSC in a system with an even number of FSs, thereby circumventing the conventional odd-FS criterion for helical TSCs.

\emph{Discussion and conclusion.}---We have introduced a novel mechanism for realizing helical topological crystalline superconductors with multiple Majorana Kramers pairs, driven by the interplay between odd-parity pairing and higher-order RSOC in locally noncentrosymmetric systems. The unique triple-winding spin texture inherent to cubic RSOC yields a TSC with a high MCN ${\cal N}_M\!\!=\!\!3$, protecting three pairs of helical Majorana edge modes. Furthermore, we have revealed a remarkable manifestation of $\bm{k}$-space engineering, where the coexistence of linear and cubic RSOCs allows both helical $p$-wave and $f$-wave pairing to emerge simultaneously on two distinct FSs.  This leads to a hybrid $p+f$-wave TSC with ${\cal N}_M\!\!=\!\!4$, a state that is topologically non-trivial despite having an even number of FSs, thus circumventing the long-standing odd-FS criterion. 

Our bilayer RSOC model is directly applicable to a range of tunable, locally noncentrosymmetric materials. Oxide heterostructures, particularly those based on LAO/STO, are prime candidates. In these systems, both linear and cubic RSOCs have been identified, and their relative strengths can be controlled by carrier filling via an external gate voltage. The experimental realization of bilayer 2D electron gases in LAO/STO superlattices~\cite{bi2DEGEXP} provides a direct structural analogue to our theoretical model, suggesting a clear path toward experimental verification. Strained germanium quantum wells, which host a high-mobility 2D hole gas with strong and complex spin-orbit effects, represent another promising platform, especially given recent progress in inducing hard-gap superconductivity in this material system. Alternatively, an effective odd-parity state can be engineered by proximitizing cubic Rashba layers with $s$-wave superconductors arranged with a $\pi$-phase difference, analogous to those proposed in other locally noncentrosymmetric multilayers~\cite{yoshida2015topological,secondvolpez}. 

While we have primarily focused on $\Delta_3$, the model also supports interlayer pairing $\Delta_2$, which can similarly give rise to a helical $f$-wave TSC. Moreover, the model admits a pairing in the 2D $E_u$ irrep ($\Delta_4$ in Table~\ref{tab:table1}), leading to a nematic superconductor that spontaneously breaks the crystal's rotational symmetry in the superconducting gap~\cite{Supp}. Although not the leading instability in our analysis, the fact that cubic RSOC naturally supports such a channel underscores its potential for stabilizing a wide range of exotic superconducting phases. The ability to generate and potentially manipulate multiple, distinct Majorana channels (${\cal N}_M\!\!=\!\!3, 4$) opens new possibilities for topological quantum information processing, moving beyond the limitations of single-qubit operations based on a single Majorana Kramers pair and enabling the exploration of more complex, multi-channel device architectures.

\emph{Acknowledgments.}---We thank Da-Shuai Ma, Zhongbo Yan, and Noah F.Q. Yuan for helpful discussions. This work was supported by National Key Research and Development Program of the Ministry of Science and Technology of China (Grant No. 2025YFA1411303), the National Natural Science Foundation of China (NSFC, Grants No. 92565103, No. 12474151, No. 12222402, and 12547101), the Natural Science Foundation of Chongqing (Grant No. CSTB2025NSCQ-LZX0010), Beijing National Laboratory for Condensed Matter Physics (No. 2024BNLCMPKF025) and the Fundamental Research Funds for the Central Universities (Grant No. 2025CDJIAISYB-032).

\let\mainbibsavedaddcontentsline\addcontentsline
\renewcommand{\addcontentsline}[3]{}
\let\addcontentsline\mainbibsavedaddcontentsline
\clearpage
\onecolumngrid
\hypersetup{pageanchor=false}
\setcounter{page}{1}
\setcounter{section}{0}
\setcounter{subsection}{0}
\setcounter{figure}{0}
\setcounter{table}{0}
\setcounter{equation}{0}
\setcounter{footnote}{0}
\makeatletter
\@ifundefined{theHsection}{}{\renewcommand{\theHsection}{SM.\arabic{section}}}
\@ifundefined{theHsubsection}{}{\renewcommand{\theHsubsection}{SM.\arabic{section}.\arabic{subsection}}}
\@ifundefined{theHfigure}{}{\renewcommand{\theHfigure}{SM.\arabic{figure}}}
\@ifundefined{theHtable}{}{\renewcommand{\theHtable}{SM.\arabic{table}}}
\@ifundefined{theHequation}{}{\renewcommand{\theHequation}{SM.\arabic{equation}}}
\@ifundefined{theHfootnote}{}{\renewcommand{\theHfootnote}{SM.\arabic{footnote}}}
\@ifundefined{theHfrontmatter}{}{\renewcommand{\theHfrontmatter}{SM.\arabic{frontmatter}}}
\makeatother

\title{Supplemental Material to ``Engineering Helical Superconductors with Multiple Majorana Kramers Pairs via Higher-Order Rashba
Spin-Orbit Coupling"}
\author{Qi-Sheng Xu}
\thanks{These authors contributed equally to this work.}
\affiliation{Department of Physics and Chongqing Key Laboratory for Strongly Coupled Physics, Chongqing University, Chongqing 400044,  China}
\author{Zi-Ming Wang}
\thanks{These authors contributed equally to this work.}
\affiliation{Department of Physics and Chongqing Key Laboratory for Strongly Coupled Physics, Chongqing University, Chongqing 400044,  China}
\author{Chui-Zhen Chen}
\affiliation{Institute for Advanced Study and School of Physical Science and Technology, Soochow University, Suzhou 215006, China}	
\author{Lun-Hui Hu}
\email{hu.lunhui.zju@gmail.com}
\affiliation{Center for Correlated Matter and School of Physics, Zhejiang University, Hangzhou 310058, China}

\author{Rui Wang}
\affiliation{Department of Physics and Chongqing Key Laboratory for Strongly Coupled Physics, Chongqing University, Chongqing 400044, China}
\affiliation{Center of Quantum Materials and Devices, Chongqing University, Chongqing 400044, China}

\author{Dong-Hui Xu}
\email{donghuixu@cqu.edu.cn}
\affiliation{Department of Physics and Chongqing Key Laboratory for Strongly Coupled Physics, Chongqing University, Chongqing 400044, China}
\affiliation{Center of Quantum Materials and Devices, Chongqing University, Chongqing 400044, China}

\maketitle

\begin{center}
{\bfseries\large Supplemental Material to ``Engineering Helical Superconductors with Multiple Majorana Kramers Pairs via Higher-Order Rashba
Spin-Orbit Coupling"}
\end{center}

\tableofcontents

\section{Bilayer Rashba Spin-Orbit Coupled Model and Symmetries}
\label{SM:Nomral state}

We begin by establishing the theoretical framework for a bilayer two-dimensional (2D) electron gas described by a low-energy continuum model expanded around the Brillouin zone center (the $\Gamma$-point, $\bm{k}=0$). This model is constructed to be consistent with the symmetries of the $D_{4h}$ point group, which is relevant for heterostructures based on materials with a square lattice structure, such as certain oxide interfaces. A key feature of the systems under consideration is the presence of local inversion symmetry breaking. While the overall bilayer structure is centrosymmetric and invariant under a global inversion operation, the individual layers lack an inversion center. This physical arrangement is designed to capture the essential physics of symmetric heterostructures, such as LaAlO$_3$/SrTiO$_3$/LaAlO$_3$, or certain van der Waals bilayers. In these systems, each layer experiences a potential gradient from its neighbor, giving rise to a Rashba spin-orbit coupling (RSOC) effect. Due to the global inversion symmetry of the bilayer, the RSOC must have opposite signs in the two layers, a configuration referred to as ``staggered" or ``layered opposite" RSOC. This local noncentrosymmetric environment is the crucial ingredient that permits the existence of RSOC, an effect typically absent in globally centrosymmetric materials.

\begin{figure}[h]
\centering
\includegraphics[width=0.96\textwidth]{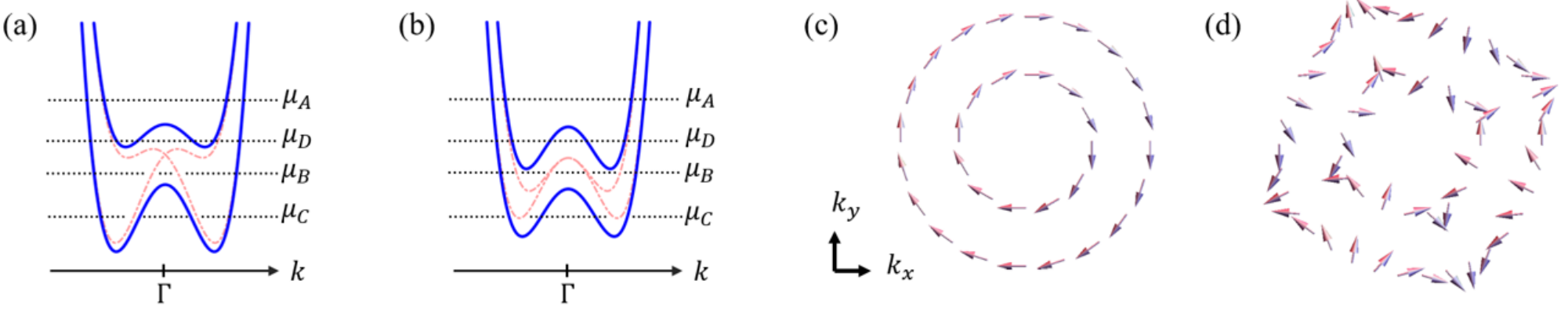}
\caption{(a), (b) Energy dispersion relations for systems with~(blue solid) and without~(red dashed) interlayer coupling for pure linear and pure cubic RSOCs, respectively. The black dashed lines indicate different positions of the chemical potential, corresponding to distinct Fermi surface topologies. (c), (d) Layer spin polarization projection onto the $xy$-plane for two distinct values of $\mu_B$ in the presence of pure linear RSOC or cubic RSOC, respectively. The arrows indicate the in-plane spin direction at a given momentum $\bm k$. The winding numbers of the spin texture are $1$ and $3$, respectively.} 
\label{SM:sfig1}
\end{figure}

The effective Hamiltonian for the normal state, ${\cal H}_{\text{N}}$, is expressed in a four-component spinor basis corresponding to the spin and layer degrees of freedom. The Hamiltonian takes the form

\begin{align}
{\cal H}_{\text{N}}(\bm{k})
&= \epsilon_0(\bm{k}) \tau_0 \otimes s_0 + \tau_z \otimes [ \bm{g}(\bm{k}) \cdot \bm{s} ] + \varepsilon \tau_x \otimes s_0  ,    
\label{SM:normalH}
\end{align} 
where the Pauli matrices $\tau_{x,y,z}$ and $s_{x,y,z}$ act on the layer {lower ($l$), upper ($u$)} and spin {$\uparrow$, $\downarrow$} degrees of freedom, respectively, and $\tau_0$ and $s_0$ are the corresponding $2\times2$ identity matrices. The first term, $\epsilon_0(\bm{k})=\beta_0 {k}^4+\beta_1 {k}^2$, describes the intra-layer kinetic energy, where the quartic term  is essential for accurately modeling the band structure away from the $\Gamma$-point and capturing the multiple Fermi surface topologies. The second term represents the staggered RSOC, where $\tau_z$ ensures the RSOC has opposite signs in the two layers, consistent with the physical picture of a globally centrosymmetric system with local inversion breaking. The RSOC vector $\bm{g}(\bm{k})=\bm{g}_{\text{LR}}(\bm{k})+\bm{g}_{\text{CR}}(\bm{k})$ includes the lowest-order terms in momentum ${\bm k}$ that are consistent with $D_{4h}$ symmetry. These are linear and cubic Rashba terms: $\bm{g}_{\text{LR}}(\bm{k})=\alpha_\text{LR}(-{k}_y,{k}_x,0)$ and $\bm{g}_{\text{CR}}(\bm{k})=\alpha_\text{CR} [-{k}_y(3{k}_x^2-{k}_y^2),{k}_x(3{k}_y^2-{k}_x^2),0]$.  The third term, with strength $\varepsilon$, is the momentum-independent interlayer tunneling, which hybridizes the states in the two layers and opens a gap of magnitude $2\varepsilon$ around the $\Gamma$ point~[Figs.~\ref{SM:sfig1}(a) and \ref{SM:sfig1}(b)]. 

The Hamiltonian in Eq. (\ref{SM:normalH}) is constructed to be invariant under the symmetry operations of the $D_{4h}$ point group. The Hamiltonian preserves time-reversal (${\cal T}$), spatial inversion (${\cal I}$), two-fold rotation (${\cal R}_{2x}$, ${\cal R}_{2y}$, ${\cal R}_{2z}$), four-fold rotation (${\cal R}_{4z}$), and mirror symmetries (${\cal M}_x$, ${\cal M}_y$, ${\cal M}_z$). The key symmetry generators and their representations in the chosen four-component basis are:
\begin{equation}
\begin{aligned}
  & {\cal T}  =i \tau_0 s_y {\cal K}, \quad {\cal I} = \tau_x s_0, \quad {\cal R}_{4z} = i \tau_x e^{i \tfrac{\pi}{2} \tfrac{s_z}{2}}, \quad {\cal R}_{2x}=i\tau_x e^{i \pi \tfrac{s_x}{2}} , \quad {\cal R}_{2y}=i\tau_x e^{i \pi \tfrac{s_y}{2}},   \\
  & {\cal R}_{2z}={\cal R}_{2x} \cdot {\cal R}_{2y}, {\quad {\cal M}_{x} = {\cal I} \cdot {\cal R}_{2x}=is_x, \quad {\cal M}_{y}={\cal I} \cdot {\cal R}_{2y}=is_y, \quad {\cal M}_{z}={\cal I} \cdot {\cal R}_{2z}=i\tau_xs_z},
  \label{SM:normal_symm_ops}
\end{aligned}
\end{equation}
where $\cal K$ is the complex conjugation operator. 

The energy spectrum of the normal state Hamiltonian can be solved analytically. The eigenvalues are given by $E_{\pm}(\bm{k})=\epsilon_0(\bm{k}) \pm \lambda_{\bm{k}}$, where $\epsilon_0(\bm{k})=\beta_0 k^4+\beta_1 k^2$ and $\lambda_{\bm{k}}=\sqrt{\varepsilon^2+|{\bm g}({\bm k})|^2}$. The Hamiltonian splits the bands into two pairs, each doubly degenerate due to the combined action of time-reversal and inversion symmetry ($\cal I \cal T$ symmetry). The interlayer coupling $\varepsilon$ opens a hybridization gap of size $2\varepsilon$ at the $\Gamma$ point. The band structures in the cases of pure linear RSOC ($\alpha_\text{CR}=0$) and pure cubic RSOC ($\alpha_\text{LR}=0$) are shown in Figs.~\ref{SM:sfig1}(a) and \ref{SM:sfig1}(b), respectively. By tuning the chemical potential, $\mu$ (indicated by the black dashed lines), the system can access different regimes, each with a distinct Fermi surface topology. These include two large, separated pockets at high doping ($\mu_A$); a single pocket within the hybridization gap ($\mu_B$); two concentric pockets ($\mu_C$), one around the $\Gamma$-point and another further away; and a final state featuring the two large, separated pockets coexisting with an additional small pocket around the $\Gamma$-point ($\mu_D$). These different topologies are crucial for determining the stability of various superconducting phases.

The spin-orbit coupling endows the electronic states with a non-trivial spin texture. A profound difference between the linear and cubic RSOC terms becomes apparent when examining the layer-dependent spin polarization, defined as $\bm p_{\eta}(\bm{k})= \langle \Psi_{\eta}(\bm{k}) | \frac{1}{2}(\tau_0+\tau_z) \bm s | \Psi_{\eta}(\bm{k}) \rangle$ for an eigenstate $\Psi_\eta(\bm k)$ of the band $E_\eta({\bm k})$, where $\eta=\pm$. The resulting spin textures of Fermi surfaces for two different values of $\mu_B$ are plotted in Figs.~\ref{SM:sfig1}(c) and \ref{SM:sfig1}(d) for the linear and cubic Rashba cases. For linear RSOC, the spin vector winds once around the origin as the momentum vector $\bm k$ completes one circuit around the Fermi surface, representing the classic spin-momentum locking with a winding number of $1$ [Fig. \ref{SM:sfig1}(c)]. In stark contrast, for cubic RSOC, the spin vector completes three full rotations for a single rotation of $\bm $ [Fig.~\ref{SM:sfig1}(d)]. This ``triple winding" texture, with a winding number of $3$, is a unique hallmark of cubic RSOC. This distinct topological feature of the normal state Fermi surface is the fundamental origin of the exotic topological superconductivity discussed later. Specifically, the triple-winding spin texture is directly responsible for the emergence of an effective helical $f$-wave pairing and the corresponding three Kramers pairs of Majorana edge modes when the system enters an odd-parity superconducting state.

\begin{table*}[!ht]
    \caption{\label{SM:stable1}%
   Classification of the six basis functions for $s$-wave orbital pairing according to the irreducible representations (irreps) of the $D_{4h}$. The pairing operators are constructed from fermion creation operators. The table also lists the parity eigenvalues under spatial inversion ({$\cal I$}) and mirror reflection about the $xy$-plane (${\cal M}_{z}$).}
   \begin{ruledtabular}
    \tabcolsep=0.1cm
    \renewcommand{\arraystretch}{1}
    \begin{tabular}{cccccc}
    Pairing & Operator representation & Matrix & Irrep. ($D_{4h}$) & ${\cal I}$ & ${\cal M}_z$ \\
    \hline
    $\Delta_1$ & $c_{l,\uparrow}c_{l,\downarrow}+c_{u,\uparrow}c_{u,\downarrow},\;\;c_{l,\uparrow}c_{u,\downarrow}-c_{l,\downarrow}c_{u,\uparrow}$ & {$i\tau_0 s_y$,\;\;$i\tau_x s_y$} & $A_{1g}$ & $+$&  $+$ \\
    $\Delta_2$ & $i${$( c_{l,\uparrow}c_{u,\downarrow}+c_{l,\downarrow}c_{u,\uparrow}$)}& {$\tau_y s_x $} & $A_{1u}$ & $-$ & $-$ \\
    $\Delta_3$ & {$c_{l,\uparrow}c_{l,\downarrow}-c_{u,\uparrow}c_{u,\downarrow}$} & {$i\tau_z s_y$} & $A_{2u}$ & $-$ & $-$ \\
    $\Delta_4$ & {$(i(c_{l,\uparrow}c_{u,\uparrow}+c_{l,\downarrow}c_{u,\downarrow}),c_{l,\uparrow}c_{u,\uparrow}-c_{l,\downarrow}c_{u,\downarrow})$}  & {($\tau_y s_z, i\tau_y s_0$)} & $E_{u}$ & $-$ &$+$
    \end{tabular}
    \end{ruledtabular}
\end{table*}

\section{Interacting Hamiltonian and Symmetry Classification of Pairing Channels}
\label{SM:SC pairs}

To investigate the superconducting instabilities, we introduce an on-site, short-range, density-density interaction. This form is a simplified model for electron-electron interactions mediated by, for example, phonons or screened Coulomb repulsion. The interaction Hamiltonian is given by
\begin{align}
    {\cal H}_\text{Int}(\bm{r}) = -U[n_{l}^2(\bm{r})+n_{u}^2(\bm{r})]-2Vn_{l}(\bm{r})n_{u}(\bm{r}),
 \label{SM:Hubburd H}
\end{align}
where $n_{\tau={l,u}}(\bm{r})\!\!=\!\!\sum_{s=\uparrow, \downarrow  }c^\dagger_{\tau,s}(\bm{r})c_{\tau,s}(\bm{r})$ denotes the electron density in the lower ($l$) and upper ($u$) layers. The parameters $U$ and $V$ represent the strengths of the intra-layer and inter-layer interactions, respectively. Positive values of $U$ or $V$ correspond to attractive interactions, which are necessary to drive superconductivity.

To ground our phenomenological interaction parameters $U$ and $V$ in physical reality, we consider the case of oxide interfaces such as LaAlO$_3$/SrTiO$_3$ (LAO/STO). In these systems, the superconducting pairing mechanism is widely understood to arise from the interplay between screened Coulomb repulsion and phonon-mediated attraction~\cite{SM-gor2016phonon}. In dilute SrTiO$3$, the Fermi energy is small ($10\text{--}50$ meV), and the system couples to longitudinal optical (LO) phonons ($\hbar\Omega_\text{LO} \approx 100$ meV). The extremely large static dielectric constant ($\epsilon_0 \sim 10^4$) strongly suppresses the long-range Coulomb repulsion. Due to this effective screening, the attractive phonon-mediated interaction dominates locally, leading to a net attractive potential $U > 0$. The screening of interactions between electrons in different layers (separated by $d \approx 4\text{--}8$ Å) is less effective than intralayer screening due to the spatial separation (finite $q_z$). Furthermore, the phonon-mediated attraction is typically local. Consequently, the residual Coulomb repulsion dominates the interlayer channel, resulting in a net repulsive interaction $V < 0$. Using weak-coupling BCS theory with typical experimental values for LAO/STO ($T_c \approx 300$ mK, $m^\ast \approx 2m_e$)~\cite{SM-shalom2010prlnov}, the effective pairing strength corresponds to $\lambda_\text{eff} \approx 0.1\text{--}0.15$. This yields an effective intralayer attraction $U \sim 80\text{--}120$ meV. Similarly, estimates of screened interlayer Coulomb repulsion yield $|V| \sim 100\text{--}160$ meV. These parameters are tunable via gate voltage $V_g$.

The Pauli exclusion principle for fermions requires the pairing wavefunction to be antisymmetric under particle exchange. For the local interaction model considered here, pairing is favored in the momentum-independent ($s$-wave orbital) channel. This orbital symmetry simplifies the antisymmetry constraint to the spin-layer part of the order parameter, which must satisfy $\Delta=-\Delta^T$. This requirement restricts the possible pairing channels to the six matrices that are antisymmetric in the four-component spin-layer basis: $ i\tau_0 s_y,i \tau_x s_y, i\tau_z s_y,\tau_y s_0, \tau_y s_x,\tau_y s_z$. These six possible pairing channels can be rigorously classified according to the irreducible representations (irreps) of the $D_{4h}$ point group. By examining how each pairing matrix $
\Delta$ transforms under the symmetry operations $g\in D_{4h}$ according to the rule $\Delta \rightarrow U(g)\Delta U^T(g)$, where $U(g)$ is the matrix representation of the symmetry operation, we can assign each channel to a specific irrep. For example, pairing matrix $ i\tau_0 s_y$ transforms trivially under all operations, identifying it as the totally symmetric $A_{1g}$ irrep. Performing this analysis for all six channels yields the classification summarized in Table \ref{SM:stable1}. The pairings group into four distinct symmetry channels, which we label $\Delta_1$ to $\Delta_4$. The classification naturally separates the channels by parity: $\Delta_1 (A_{1g})$ is an even-parity channel, while $\Delta_2 (A_{1u})$, $\Delta_3 (A_{2u})$ and $\Delta_4 (E_u)$ are odd-parity channels. Odd-parity superconductivity is a prerequisite for many forms of topological superconductivity. The existence of a two-dimensional irrep, $E_u$, signals the possibility of spontaneous rotational symmetry breaking, leading to nematic superconductivity, which is explored in Section~\ref{SM:nematic SC}.

\section{Superconducting Linearized Gap Equations and Phase Diagrams}
\label{SM:GL theory}

To determine the dominant superconducting instability as a function of system parameters, we derive and solve the linearized gap equations within the weak-coupling BCS framework. The superconducting transition temperature $T_c$ is found by identifying the temperature at which a non-trivial solution to these equations first appears. The analysis begins with the normal state electron and hole Green's functions in the Matsubara formalism. For the Hamiltonian ${\cal H}_{\text{N}}(\bm{k})=\epsilon_0(\bm k)\tau_0s_0+{\cal H}_1(\bm k)$, the Green's functions for the two bands $E_{\eta=\pm}$ are 
\begin{equation}
    \mathcal{G}_{e,h}(\omega,\bm{k})=\frac{P_\eta(\bm k)}{\mp i \omega_n\pm E_{\eta}(\bm{k})\mp\mu},
\end{equation}
where $\omega_n=(2n+1)\pi k_BT$ are the fermionic Matsubara frequencies, and $P_{\eta=\pm}=\frac{1}{2}(1\pm \frac{H_1(\bm{k})}{\lambda_k})$ are projection operators onto the respective energy bands. The linearized gap equation for a pairing channel
$\Delta_i$ driven by an interaction vertex $V_i$ takes the general form $\Delta_i=V_i\chi_i \Delta_i$, which has a non-trivial solution when the largest eigenvalue of the matrix $V_i\chi_i$ is $1$. The pairing susceptibility $\chi_i$ is given by the standard bubble diagram:

\begin{align}
	\chi_{i}=-\frac{1}{\beta_c}\sum_{\omega_n,\bm{k}}\mathrm{Tr}[\Gamma^\dagger_{i}\mathcal{G}_e(\omega,\bm{k})\Gamma_{i}\mathcal{G}_h(\omega,\bm{k})],
 \label{SM:seq4}
\end{align}
where $\beta_c=1/k_BT_c$, and $\Gamma_i$ is the matrix representation of the pairing operator $\Delta_i$. Since pairings belonging to different irreps do not mix, the gap equations for each channel can be treated separately. For the$ A_{1g}$ channel, which involves a mixture of two basis functions driven by interactions $U$ and $V$, the condition for a non-trivial solution is given by a determinant equation

\begin{align}
\begin{vmatrix}
 U\chi _{1a}-1 & U \chi _{1ab} \\
 V\chi _{1ba} & V\chi _{1b}-1
\end{vmatrix}=0.
\label{SM:seq5}
\end{align}
For the other channels, which are one-dimensional (or treated as such for the degenerate $E_u$ case), the conditions are simpler scalar equations:
\begin{align}
 V \chi_2=1, U \chi_3=1, V \chi_4=1.
 \label{SM:seq6}
\end{align}
After performing the Matsubara summation, the susceptibility for each channel $i$ can be expressed as an integral over the Brillouin zone:

\begin{equation}
    \begin{aligned}\chi_{i}
           &=-\frac{1}{N} \sum_{\bm{k}, \eta=\pm}\Bigg [ F_{i}^{\eta}(\bm{k})\frac{\tanh[\beta_c(E_\eta-\mu)/2]}{2(E_\eta-\mu)}\Bigg ].
    \label{SM:seq4-2} 
    \end{aligned}
\end{equation}
The form factors $F^\eta_i(\bm k)$ represent the projection of the pairing interaction onto the spin-split normal state bands. For the channels considered, they are: $F^\pm_{1a}(\bm{k})=1$, $F^+_{1ab}(\bm{k})=-F^-_{1ab}(\bm{k})=\tfrac{\varepsilon}{\lambda_k}$, $F^+_{1ba}(\bm{k})=-F^-_{1ba}(\bm{k})=\tfrac{\varepsilon}{\lambda_k}$, $F^\pm_{1b}(\bm{k})=\tfrac{\varepsilon^2}{\lambda_k^2}$, $F^\pm_{2}(\bm{k})=1-\tfrac{\varepsilon^2}{\lambda_k^2}$, $F^\pm_{3}(\bm{k})=1-\tfrac{\varepsilon^2}{\lambda_k^2}$, $F^\pm_{4}(\bm{k})=\frac{1}{2}(1-\tfrac{\varepsilon^2}{\lambda_k^2})$.

\begin{figure}
\centering
\includegraphics[width=0.49\textwidth]{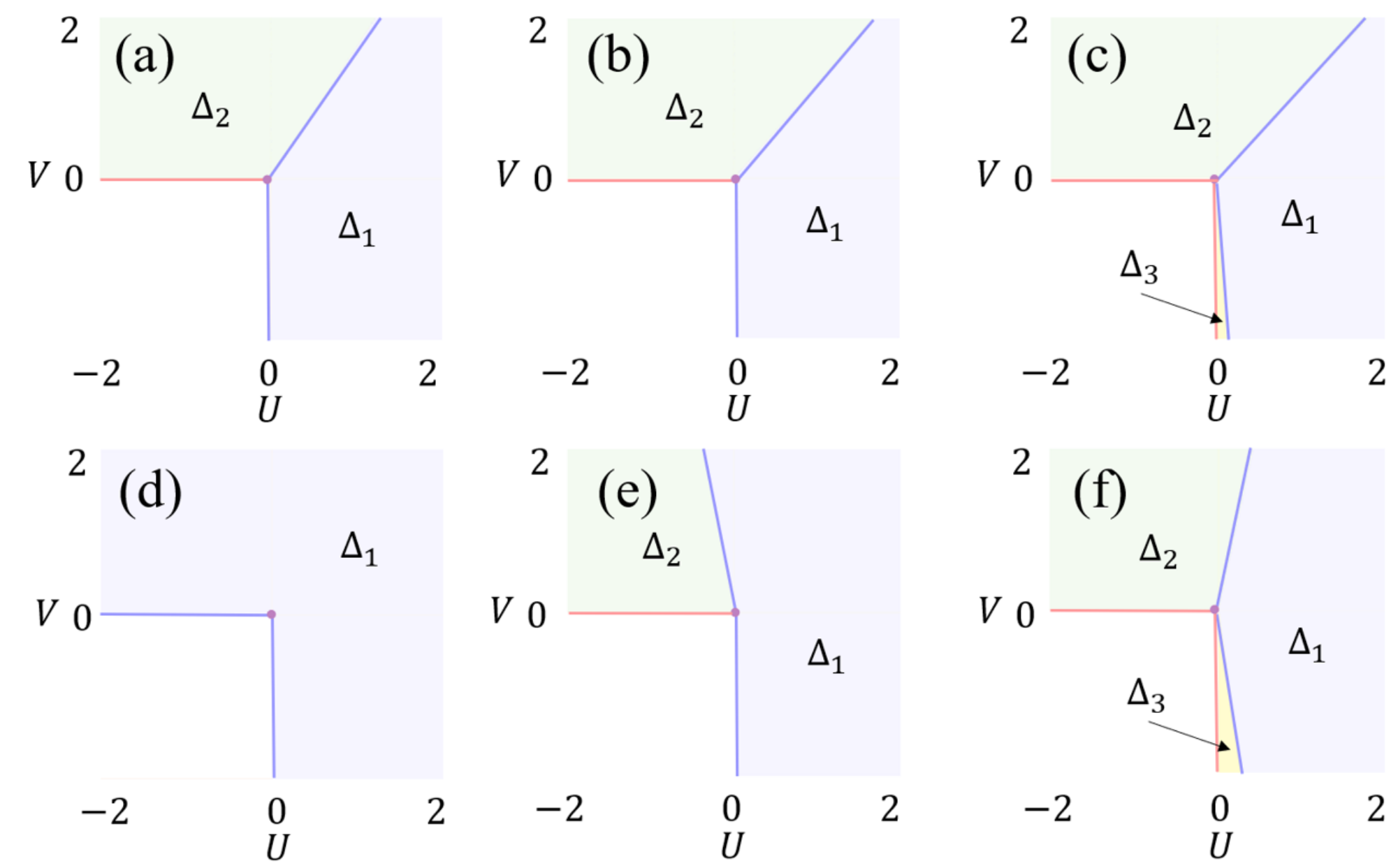}
\includegraphics[width=0.49\textwidth]{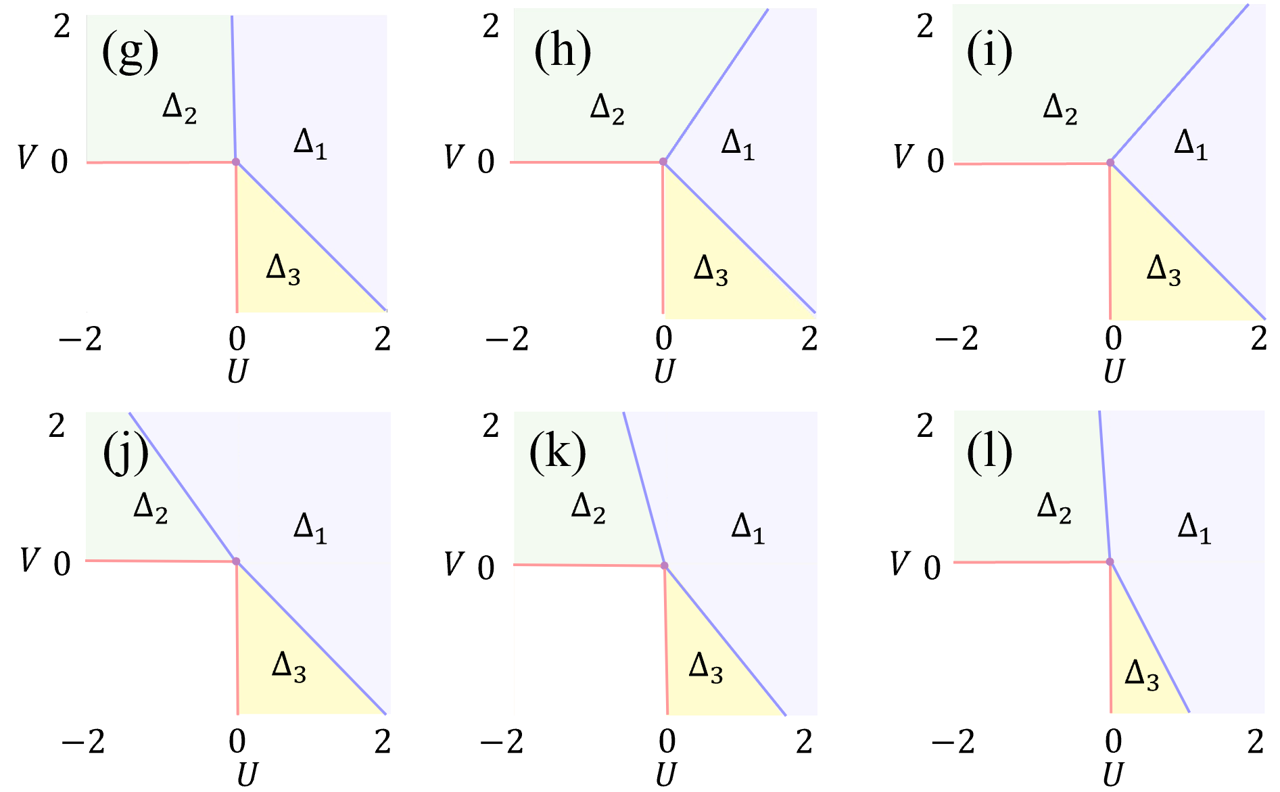}
\caption{Phase diagrams of the leading superconducting pairing channel in the plane of inter-layer ($V$) and intra-layer ($U$) interaction strengths for a system with increasing pure cubic RSOC strength. The chemical potential values are chosen as $\mu_A=1.05$, $\mu_D=0.15$, $\mu_B=-0.2$, and $\mu_C=-0.65$, corresponding to the four representative cases as shown in Fig.~\ref{SM:sfig1} (b). (a), (b), and (c) represent the phase diagram in case of $\mu_A$ when $\alpha_{\text{CR}}$ is $0.1$, $0.2$, and $0.3$, respectively. (d), (e), and (f) represent the phase diagram in case of $\mu_D$ when $\alpha_{\text{CR}}$ is $0.1$, $0.2$, and $0.4$, respectively. (g), (h) and (i) represent the phase diagram in case of $\mu_B$ when $\alpha_{\text{CR}}$ is $0.1$, $0.2$, and $0.3$, respectively. (j), (k), and (l) represent the phase diagram in case of $\mu_C$ when $\alpha_{\text{CR}}$ is $0.1$, $0.2$, and $0.3$, respectively.The green region represents the $\Delta_1$ phase, the yellow region represents the $\Delta_2$ phase, the red region represents the $\Delta_3$ phase, and the white region represents the non-superconducting phase. Common parameters, $\beta_0=0.5,\beta_1=-1,\varepsilon=0.35$.}
\label{SM:sfig2}
\end{figure}

The channel with the highest critical temperature $T_c$ corresponds to the leading superconducting instability. By numerically solving these equations for the critical temperature for each channel, we determine the leading superconducting instability and construct the phase diagrams shown in Fig.~\ref{SM:sfig2}. These diagrams are calculated for four different Fermi surface topologies ($\mu_A$-$\mu_D$) while varying strengths of the cubic RSOC, $\alpha_\text{CR}$. These diagrams reveal a rich competition between pairing channels, which depends sensitively on the Fermi surface topology and the strength of the cubic RSOC. We analyze the evolution of the phase diagram as $\alpha_\text{CR}$ increases. In case of $\mu_A$ [Figs.~\ref{SM:sfig2}(a)-\ref{SM:sfig2}(c)], when the attractive intralayer interaction ($U>0$) is dominant, the system favors the even-parity $\Delta_1$ phase. The odd-parity $\Delta_2$ phase is favored when the attractive interlayer interaction ($V>0$) becomes dominant to lead to superconducting instability. At a sufficiently large $\alpha_\text{CR}$, the odd-parity $\Delta_3$ phase emerge in a small area. In case of $\mu_D$ [Figs.~\ref{SM:sfig2}(d)-\ref{SM:sfig2}(f)], at weak $\alpha_\text{CR}$, only the $\Delta_1$ phase prevails regardless of whether the intralayer or interlayer attraction is dominant. When $\alpha_\text{CR}$ increase, the odd-parity $\Delta_2$ and $\Delta_3$ phases emerge, although the $\Delta_3$ phase still occupies a small area. Finally, in cases of $\mu_B$ and $\mu_C$ where only lower band is partially filled [Figs.~\ref{SM:sfig2}(g)-\ref{SM:sfig2}(l)], the three phases are competitive even at weak $\alpha_\text{CR}$. As $\alpha_\text{CR}$ increases, the regions dominated by odd-parity phases generally expand. 

The intralayer odd-parity $\Delta_3$ is of primary focus of this work. Its stability region appears in cases $\mu_B$ and $\mu_C$. The emergence of this odd-parity state as a ground state is plausible in realistic material settings, such as oxide interfaces, where superconductivity arises from a competition between screened Coulomb repulsion and phonon-mediated attraction. The parameters $U$ and $V$ represent the net effective interactions resulting from this interplay. It is reasonable to assume that both screening and phonon coupling are more effective within the layers than between them. This asymmetry can create a scenario where the effective intralayer interaction $U$ becomes attractive ($U>0$) while the effective interlayer interaction $V$ remains repulsive ($V<0$). Therefore, we anticipate that the $\Delta_3$ superconducting pairing is a realistic possibility in bilayer cubic Rashba systems.

\begin{figure}[h]
\centering
\includegraphics[width=0.49\textwidth]{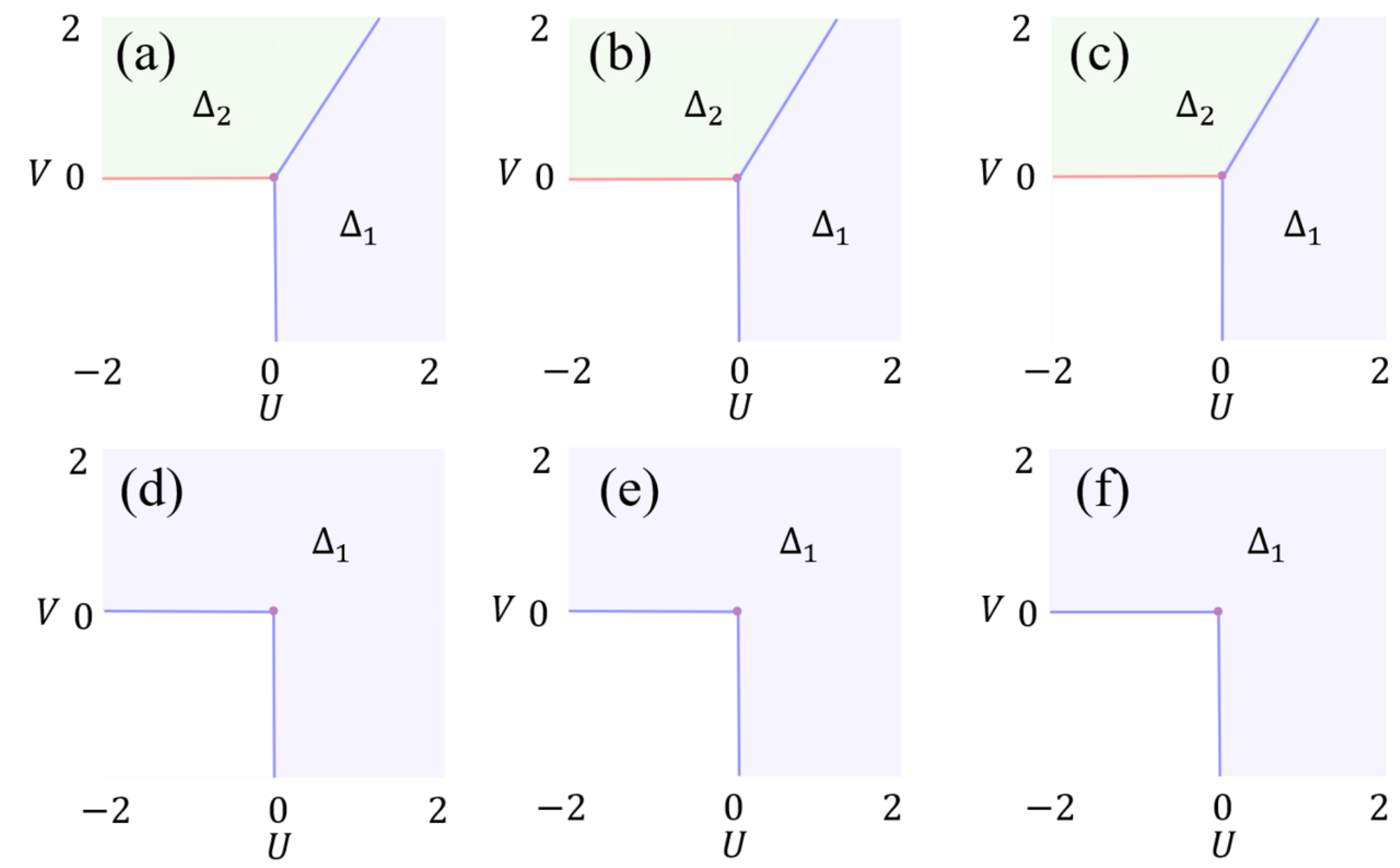}
\includegraphics[width=0.49\textwidth]{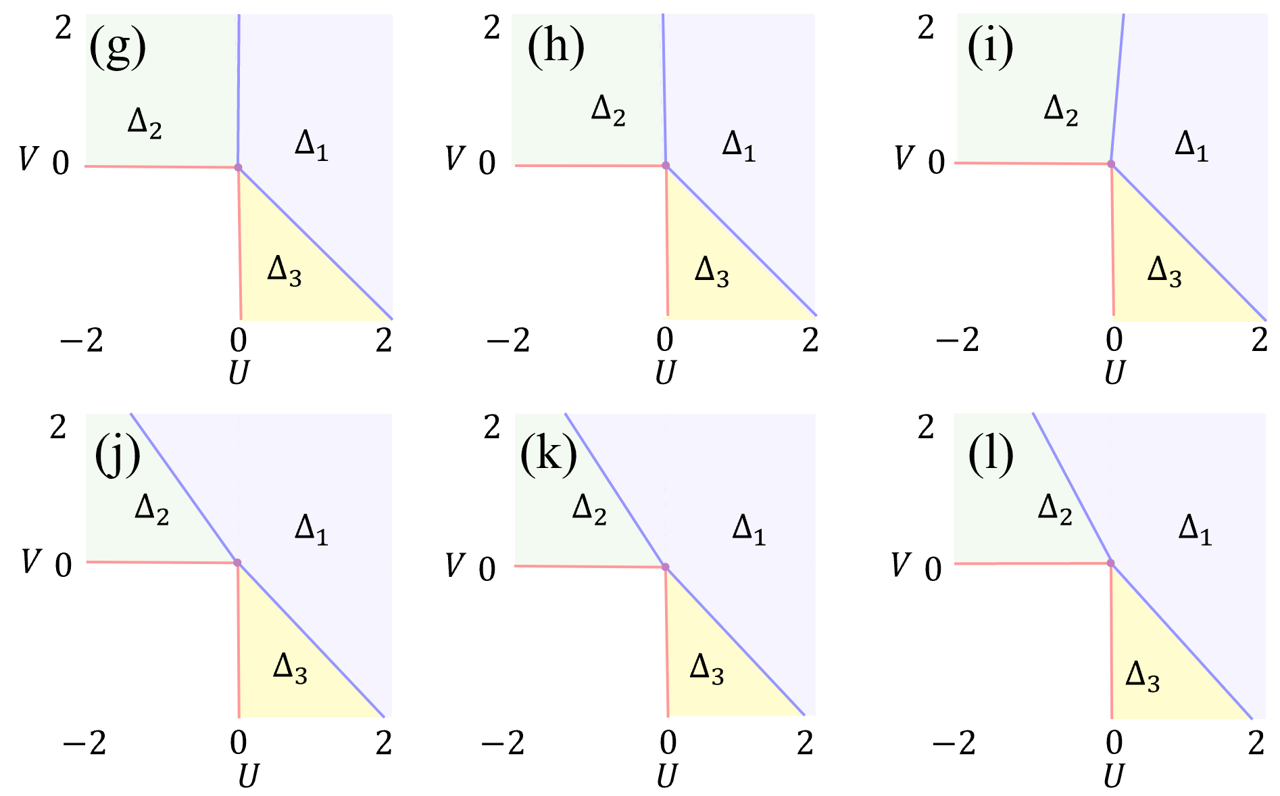}
\caption{Superconducting phase diagrams for different pairing channels. Evolution of superconducting phase diagrams with $\alpha_\text{LR}$ for different pairing channels and chemical potentials $\mu$. Columns (left to right) correspond to $\alpha_\text{LR} = 0.05, 0.1, 0.2$. Rows (top to bottom) correspond to: (a)-(c) $\mu_A$=1.05, (d)-(f) $\mu_D$=0.15, (g)-(i) $\mu_B$=-0.2, (j)-(l) $\mu_C$=-0.65. Other parameters are fixed:  $\beta_0=0.5,\beta_1=-1,\varepsilon=0.35,\alpha_\text{CR} = 0.1$}
\label{SM:sfig3}
\end{figure}

To more closely approximate real materials, we further investigate the effects of coexisting linear Rashba spin-orbit coupling (RSOC). Figure~\ref{SM:sfig3} shows the resulting phase diagrams when a finite linear RSOC ($\alpha_{\text{LR}}$) is included. These phase diagrams are calculated for various strengths of the linear RSOC, with the cubic RSOC kept fixed. For both $\mu_A$ and $\mu_D$ cases [Figs.~\ref{SM:sfig3}(a)-\ref{SM:sfig3}(f)] the phase diagrams exhibit negligible changes. In the case of $\mu_B$ [Figs.~\ref{SM:sfig3}(g)-\ref{SM:sfig3}(i)], as the linear RSOC increases, the region of the $\Delta_3$ phase remains unchanged, while that of the $\Delta_2$ phase shifts gradually to the right. Finally, for $\mu_C$ [Figs.~\ref{SM:sfig3}(j)-\ref{SM:sfig3}(l)], the behavior of $\Delta_2$ resembles that in $\mu_B$ but with a smaller shift. Additionally, the phase boundary between $\Delta_3$ and $\Delta_1$ undergoes a slight displacement in the fourth quadrant, resulting in a modest reduction of the $\Delta_3$ phase region. Although the phase boundaries shift with increasing $\alpha_{\text{LR}}$, the $\Delta_3$ phase of interest persists across a significant portion of the phase diagram, particularly in the physically relevant regime (attractive $U$, repulsive $V$) in cases of $\mu_B$ and $\mu_C$.

We justify the use of the BdG mean-field formalism for this quasi-2D system based on two main criteria: I. The ratio of the superconducting gap to the Fermi energy is $\Delta/E_F \sim (0.05 \text{ meV}) / (20 \text{ meV}) \sim 10^{-3}$. In this weak-coupling limit, the mean-field ground state provides an accurate description of the Cooper instability and the resulting quasiparticle spectrum. II. We acknowledge that in 2D, phase fluctuations suppress long-range order at finite temperature ($T_\text{BKT} < T_\text{MF}$). However, $T_\text{MF}$ correctly identifies the pairing energy scale $T^\ast$. Since the topological invariants  and Majorana edge modes are properties of the gapped ground state ($T=0$), they are robustly captured by the BdG mean-field formalism.

Our analysis targets doped regimes without special nesting. In weak coupling, the Cooper channel provides the dominant logarithmic instability; density-wave instabilities typically require strong nesting and/or strong repulsion. For the approximately circular Fermi surfaces and strongly screened interactions expected in gate-tuned interface regimes, superconductivity is therefore the leading instability within the model's intended parameter window.

\section{Topological Superconductor Phases}
\label{SM:TSC phase}
Among the competing superconducting phases identified, the odd-parity $\Delta_3$ ($A_{2u}$) phase is of particular interest as it gives rise to a topological superconductor (TSC) characterized by multiple helical Majorana edge modes and an effective helical $f$-wave pairing symmetry.

\subsection{Helical $f$-Wave Superconductivity}
\label{SM:fwave}
We now analyze the topological properties of the superconducting phase driven by the order parameter $\Delta_3$. The normal state respects the mirror symmetry ${\cal M}_z=i\tau_x s_z$ with respect to the $xy$-plane. The $\Delta_3$  order parameter is odd under this operation, leading to a superconducting state that spontaneously breaks ${\cal M}_z$. The BdG Hamiltonian retains a modified mirror symmetry by introducing operator $\tilde{\cal M}_z\equiv\text{diag}[{\cal M}_z, -{\cal M}^{\ast}_z]$ in the Nambu space. This extended symmetry allows the BdG Hamiltonian to be block-diagonalized into two sectors, corresponding to the mirror eigenvalues $\pm i$. The BdG Hamiltonian matrix for each sector is ${\cal H}_{\pm i}(\bm{k})$, where 
\begin{align*}
{\cal H}_{\pm i}(\bm{k})=\begin{pmatrix}
 {\cal H}_{\text{N},\pm i}(\bm{k}) &-i\Delta_0 s_y  \\
 i \Delta_0 s_y  & -{\cal H}^\ast_{\text{N},\pm i}(-\bm{k}) 
\end{pmatrix}.
\end{align*}
${\cal H}_{+i}(\bm{k})$ and ${\cal H}_{-i}(\bm{k})$ are connected by TRS, leading to the absence of time-reversal symmetry within each block individually.
To reveal the effective pairing of the superconductor, it is necessary to transform the Hamiltonian into the band basis. The normal state counterpart of block Hamiltonian ${\cal H}_{\text{N},+i}(\bm{k}) $ can be diagonalized by using the following expression
 \begin{equation}
\label{SM:Utrans}
\begin{aligned}
       \Psi(\bm{k})=\Phi_+(\bm{k})\Psi_+(\bm{k})+\Phi_-(\bm{k})\Psi_-(\bm{k}),
\end{aligned}
\end{equation}
where $\Psi_+$ ($\Psi_-$) represents the annihilation operator for the upper (lower) energy band, and $\Phi_+$ ($\Phi_-$) denotes the corresponding normalized wavefunction
\begin{equation}
\begin{aligned}
&\Phi_{+}(\bm{k})=\sqrt{\frac{1}{2\lambda_k(\lambda_k+\varepsilon)}}\begin{pmatrix}\lambda_k+\varepsilon\\ g_x+ig_y\end{pmatrix},\\
&\Phi_{-}(\bm{k})=\sqrt{\frac{1}{2\lambda_k(\lambda_k+\varepsilon)}}\begin{pmatrix}-g_x+ig_y\\\lambda_k+\varepsilon\end{pmatrix},
\end{aligned}
\end{equation}
where $g_{x,y}$ are the $x$ and $y$ components of $\bm{g}(\bm{k})$, respectively. Therefore, the block BdG Hamiltonian $H_{+i}(\bm{k})$ in the band basis reads 
\begin{equation}
\label{SM:projected}
\begin{aligned}
H_{+i}(\bm{k})
&=\xi_+(\bm{k})\Psi_+^\dagger(\bm{k})\Psi_+(\bm{k})+\xi_-(\bm{k})\Psi_-^\dagger(\bm{k})\Psi_-(\bm{k})\\&+\Delta_{++}(\bm{k})\Psi_+^\dagger(\bm{k})\Psi_+^\dagger(-\bm{k})+\Delta_{+-}(\bm{k})\Psi_+^\dagger(\bm{k})\Psi_-^\dagger(-\bm{k})\\ &+\Delta_{--}(\bm{k})\Psi_-^\dagger(\bm{k})\Psi_-^\dagger(-\bm{k})+\text{H.C.},
\end{aligned}
\end{equation}
with
\begin{equation*}
\begin{aligned}
&\xi_{\pm}(k)=-\epsilon_0(\bm{k}) \pm \lambda_{\bm k},\\
&\Delta_{++}(\bm{k})=-\Delta_0\frac{\alpha_\text{CR}}{\lambda_{\bm k}}\Big(-ik_x+k_y\Big)^3,\\
&\Delta_{+-}(\bm{k})=-\Delta_0\frac{\varepsilon}{\lambda_{\bm k}}, \\
&\Delta_{--}(\bm{k})=-\Delta_0\frac{\alpha_\text{CR}}{\lambda_{\bm k}}\Big(ik_x+k_y\Big)^3.
\end{aligned}
\end{equation*}
Equation~\eqref{SM:projected} reveals that the pairing $\Delta_3$ not only induces the inter-band $s$-wave pairing but also the intra-band $f_x\pm i f_y$ pairing with opposite chirality for the upper and lower bands. When the chemical potential $\mu_B$ is within the hybridization gap, the Fermi surface is solely provided by the lower band. Therefore, $H_{+i}(\bm{k})$ describes a $f + if$ wave superconductor, while $H_{-i}(\bm{k})$ corresponds to the $f-if$ wave superconductor. These two sectors form a time-reversal invariant TSC with helical $f$-wave pairing.

To explicitly demonstrate this protection, we compute the mirror Chern number (MCN) ${\cal N}_{\text{M}}=({\cal N}_{+i}-{\cal N}_{-i})/2$~\cite{SM-PhysRevB.78.045426,SM-ueno2013symmetry,SM-chiu2013classification,SM-tsutsumi2013upt3,SM-shiozaki2014topology,SM-ando2015topological,SM-yoshida2015topological}, where ${\cal N}_{+i}$ and ${\cal N}_{-i}$ are the Chern numbers associated with mirror symmetry eigenvalues $+i$ and $-i$ for all occupied bands, respectively. Since each mirror subsector belongs to class D, its topology is characterized by an integer Chern number, ${\cal N}_{\pm i}$. A direct calculation shows that when the chemical potential is inside the hybridization gap and the system is fully gapped (requiring $\varepsilon^2>\mu^2+\Delta_0^2$), the MCN is ${\cal N}_{\text{M}}=3$. 

\begin{figure}[h]
\centering
\includegraphics[width=0.72\textwidth]{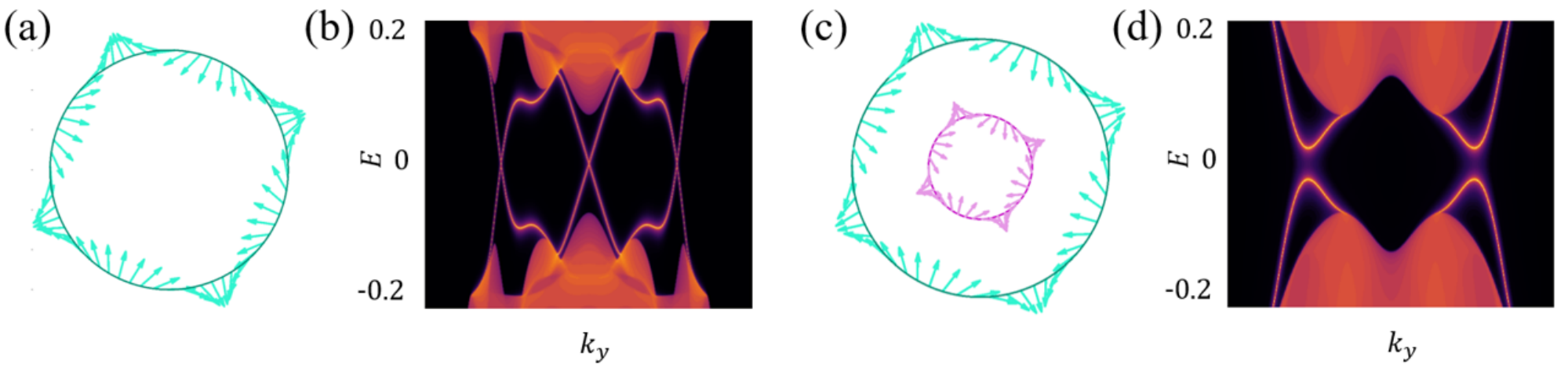}
\caption{Fermi surfaces with spin texture for $\mu$ located at different position and spectral function of $\Delta_3$ state in the presence of only cubic RSOC. (a) Single Fermi surface with spin texture for case B, where $\mu_B=-0.2$ is located within the hybridization gap. (b) Spectral function of $\Delta_3$ states for the chemical potential in (a). (c) Two concentric Fermi surfaces with spin texture for case C, $\mu_C=-0.6$. (d) Spectral function of $\Delta_3$ states for the chemical potential in (c). Common parameters: $\beta_0=0.5,\beta_1=-1,\varepsilon=0.35,\Delta_0=0.2, \alpha_{\text{CR}}=0.1$.}
\label{SM:sfig4}
\end{figure}

 The bulk-boundary correspondence for this topological crystalline superconductor dictates that a bulk MCN of ${\cal N}_{\text{M}}=3$ guarantees the existence of $|{\cal N}_{\text{M}}|$ pairs of counter-propagating helical Majorana edge modes protected by the combination of time-reversal and mirror symmetry. To explicitly demonstrate these edge states, we discretize the continuum model on a square lattice.  By choosing the basis ${\Psi}^{\dagger}_{\bm{k}}=(c^\dagger_{u,\bm{k},\uparrow},c^\dagger_{u,\bm{k},\downarrow},{c^\dagger_{l,\bm{k},\uparrow}},c^\dagger_{l,\bm{k},\downarrow},c_{u,-\bm{k},\uparrow}, {c_{u,-\bm{k},\downarrow}},c_{l,-\bm{k},\uparrow},c_{l,-\bm{k},\downarrow})$, the Bogoliubov-de Gennes (BdG) Hamiltonian on the square lattice is given by

\begin{equation}
\begin{aligned}
    {\cal H}_{\text{BdG}} &=[t_0(\cos{2k_x}+\cos{2k_y})+t_1(\cos{k_x}+\cos{k_y})-2t_0-2t_1-\mu]\sigma_z \tau_0 s_0\\&-\alpha_{\text{CR}}(4\sin{k_y}-6\cos{k_x}\sin{k_y}+\sin{2k_y})\sigma_0\tau_z s_x\\&+\alpha_{\text{CR}}(4\sin{k_x}-6\cos{k_y}\sin{k_x}+\sin{2k_x})\sigma_z \tau_z s_y\\&+\varepsilon \sigma_z \tau_x s_0-\Delta_0 \sigma_y \tau_z s_y,
\end{aligned}
\label{SM:seq8}
\end{equation}
where $\sigma_{x,y,z}$ act in Nambu space, $t_0$ represents the strength of second-nearest neighbor hopping, $t_1$ represents the strength of nearest-neighbor hopping, $\mu$ is the chemical potential, and $\Delta_0$ denotes the pairing amplitude of odd-parity pairing $\Delta_3$. $\alpha_\text{CR}$ represents the coefficient of cubic RSOC. $\varepsilon$ denotes the interlayer coupling, which induces the hybridization gap around the $\Gamma$ point. Figure~\ref{SM:sfig4} shows the Fermi surface topologies corresponding to the cases of $\mu_B$ and $\mu_C$ and the corresponding spectral functions for a ribbon geometry with $\Delta_3$ pairing. When the chemical potential lies within the hybridization gap [Fig.~\ref{SM:sfig4}a], these are the predicted three pairs of helical Majorana edge modes [Fig.~\ref{SM:sfig4}b], a direct consequence of the ${\cal N}_{\text{M}}=3$ bulk topology. In contrast, when the chemical potential cuts an even number of Fermi surfaces enclosing the $\Gamma$-point [Fig.~\ref{SM:sfig4}c], the system is topologically trivial, and the spectral function shows gapped Majorana edge modes [Fig.~\ref{SM:sfig4}d]. 

To further assess the robustness of the helical Majorana edge spectrum, we introduced a set of symmetry-breaking perturbations (with amplitudes $\le 0.1$) that selectively violate the symmetry operations defined in Eq.~\eqref{SM:normal_symm_ops}. Representative results are summarized in Table~\ref{SM:stable2}, where ``OpenX" (``OpenY") denotes the number of helical Majorana Kramers pairs under open boundary conditions along the $x$ ($y$) direction.

As shown in Table~\ref{SM:stable2}, as long as the mirror symmetry ${\cal M}_z$ is preserved, the system robustly supports three Kramers pairs of helical Majorana edge modes, independent of the edge orientation (OpenX and OpenY both yield $3$). In contrast, once ${\cal M}_z$ is broken, two distinct situations can occur. (i) If time-reversal symmetry ${\cal T}$ remains intact but no additional crystalline protection is enforced, the extra Majorana pairs gap out and only a single Kramers pair survives, consistent with the $\mathbb{Z}_2$ classification of 2D Class DIII superconductors. (ii) For certain edge terminations, gapless edge modes can still be protected by combined symmetries such as ${\cal P}\cdot{\cal R}_{2x}$ or ${\cal P}\cdot{\cal R}_{2y}$ with ${\cal P}$ the particle-hole symmetry, leading to boundary-dependent stability of the edge spectrum. This edge-selective protection is closely related to the notion of crystalline-symmetry-pinned Majorana responses (often discussed in terms of ``Majorana Ising" behavior)~\cite{SM-shiozaki2014topology}.

Since the unperturbed phase exhibits three helical Majorana Kramers pairs for all boundary orientations, whereas the combined-symmetry protection appears only for specific edges once ${\cal M}_z$ is broken, these results support the conclusion that the three-pair helical phase is protected by the global mirror symmetry ${\cal M}_z$.

\begin{table*}[!ht]
    \caption{\label{SM:stable2}%
    Symmetry-breaking perturbations and the resulting helical Majorana edge modes.}
    \begin{ruledtabular}
        \begin{tabular}{lcccccc} % 
            \textrm{Perturbation} &
            \textrm{${\cal T}$} &
            \textrm{${\cal P}\cdot {\cal R}_{2x}$} &
            \textrm{${\cal P}\cdot {\cal R}_{2y}$} &
            \textrm{${\cal M}_z$} &
            \textrm{OpenX} &
            \textrm{OpenY} \\
            \colrule
            {$\sigma_0\tau_xs_0$} & {$\surd $} & {$\times$} & {$\times$} & {$\surd $} & {3} & {3} \\
            {$\sigma_z\tau_zs_0$} & {$\surd $} & {$\times$} & {$\times$} & {$\times$} & {1} & {1} \\
            {$\sigma_z\tau_xs_z$} & {$\times$} & {$\times$} & {$\times$} & {$\surd $} & {3} & {3} \\{$\sigma_z\tau_zs_z$} & {$\times$} & {$\surd $} & {$\surd $} & {$\times$} & {3} & {3} \\
            {$\sigma_0\tau_xs_y$} & {$\times$} & {$\times$} & {$\surd $} & {$\times$} & {0} & {3} \\
            {$\sigma_z\tau_0s_x$} & {$\times$} & {$\surd $} & {$\times$} & {$\times$} & {3} & {0} \\
            {$\sigma_0\tau_ys_0$} & {$\times$} & {$\times$} & {$\times$} & {$\times$} & {0} & {0} \\

        \end{tabular}
    \end{ruledtabular}
\end{table*}

\subsection{Hybrid Helical $p+f$-Wave Superconductivity}
\label{SM:Coexisting linear and cubic RSOC}
\begin{figure}[h]
\centering
\includegraphics[width=0.72\textwidth]{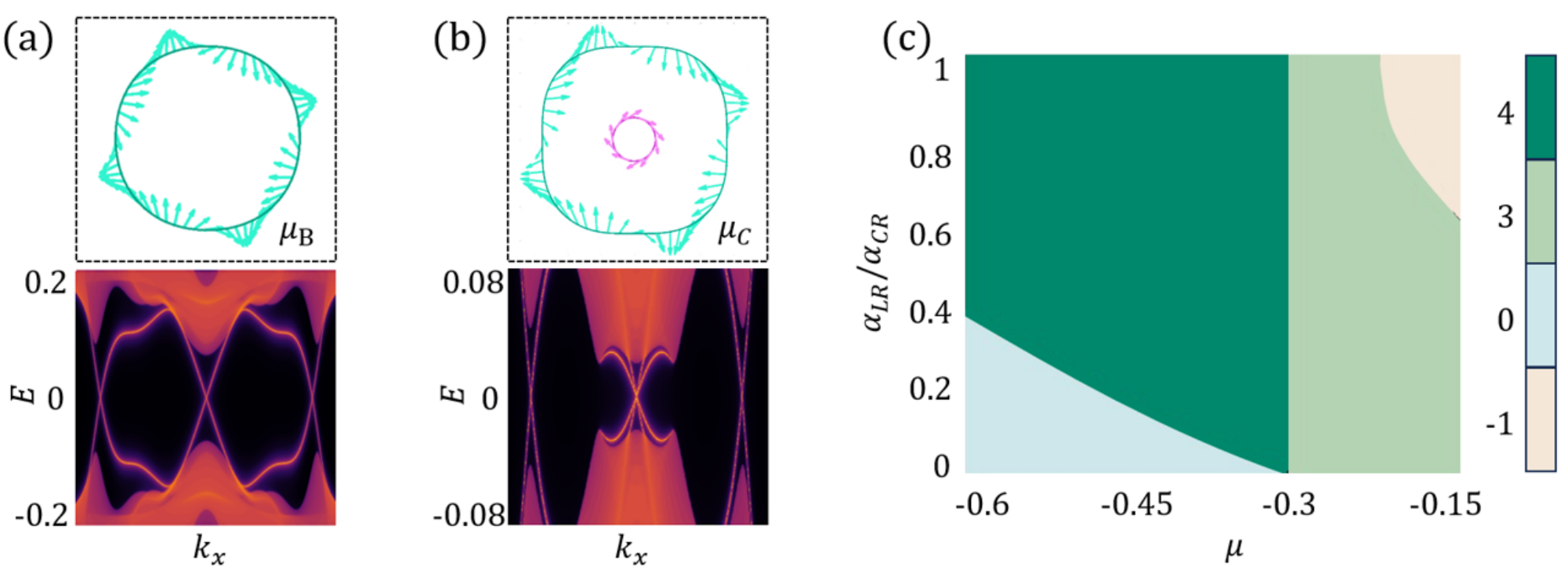}
\caption{Fermi surfaces, spin texture and spectral function of $\Delta_3$ state with coexisting linear and cubic RSOCs. (a) Single Fermi surface within the hybridization gap and its spin texture (top) for $\mu_B=-0.2, \alpha_{\text{LR}}=0.06$. The corresponding spectral function of $\Delta_3$ phase is shown below. (b) Two concentric Fermi surfaces enclosing the $\Gamma$ point (top) for $\mu_C=-0.4, \alpha_{\text{LR}}=0.1$. The corresponding spectral function of $\Delta_3$ phase is shown below. (c) Topological phase diagram determined by MCN in the regime where the chemical potential intersects only the lower band. Common parameters:  $\beta_0=0.5, \beta_1=-1, \alpha_{\text{CR}}=0.1, \varepsilon=0.35, \Delta_0=0.2$.} 
\label{SM:sfig5}
\end{figure}

In real materials, for instance, in oxide heterostructures like LaAlO$_3$/SrTiO$_3$/LaAlO$_3$~\cite{SM-lin2019interface,SM-caviglia2010tunable}, linear and cubic RSOCs  can coexist, with their relative strengths tunable via carrier filling. We therefore investigated the robustness of the ${\cal N}_{\text{M}}=3$ phase by reintroducing the linear RSOC ($\alpha_\text{LR} \neq 0$). As long as the chemical potential lies within the hybridization gap, the system exhibits a single Fermi surface enclosing the time-reversal invariant $\Gamma$-point [Top panel of Fig.~\ref{SM:sfig5}(a)]. We found that the three pairs of helical Majorana modes persist, provided the cubic RSOC remains dominant~[Bottom panel of Fig.~\ref{SM:sfig5}(a)].

Crucially, a new phenomenon emerges when the chemical potential is moved out of this gap but still cuts the lower band. This creates two distinct FSs: a smaller, inner surface near the $\Gamma$ point and a larger, outer surface farther away~[Top panel of Fig.~\ref{SM:sfig5}(b)]. In this regime $\Delta_3$ pairing remains favored under similar interaction conditions. The interplay between coexisting linear and cubic RSOCs and $\Delta_3$ pairing induces an even richer TSC characterized by a heightened MCN, ${\cal N}_{\text{M}}=4$, hosting four Kramers pairs of Majorana edge modes [Bottom panel of Fig.~\ref{SM:sfig5}(b)]. This state seemingly conflicts with the topological criteria for odd-parity superconductors~\cite{SM-Sato-odd,SM-oddparityTscFu}. However, the different momentum scaling of the two RSOCs allows them to dominate in different regions of the Brillouin zone. Given sufficient separation of the FSs in $k$-space, the linear term ($\propto k$) dominates the inner surface, fostering a helical $p$-wave state. Conversely, the cubic term ($\propto k^3$) dominates the outer surface, fostering a helical $f$-wave state. The resulting superconductor effectively merges these two distinct topologies, summing their respective contributions to the total MCN and realizing a hybrid helical $p+f$-wave pairing. This result highlights a general principle for engineering high MCN topological phases by leveraging the momentum-space separation of different physical mechanisms. The topological phase diagram in Fig.~\ref{SM:sfig5}(c) maps the stability of the ${\cal N}_{\text{M}}=-1, 3$, and $4$ which are induced by the odd-parity pairing $\Delta_3$, as a function of the chemical potential and the ratio of linear to cubic RSOC strength.

\clearpage
\section{Nematic Superconductor Phase}
\label{SM:nematic SC}
\begin{figure}[H]
\centering
\includegraphics[width=0.96\textwidth]{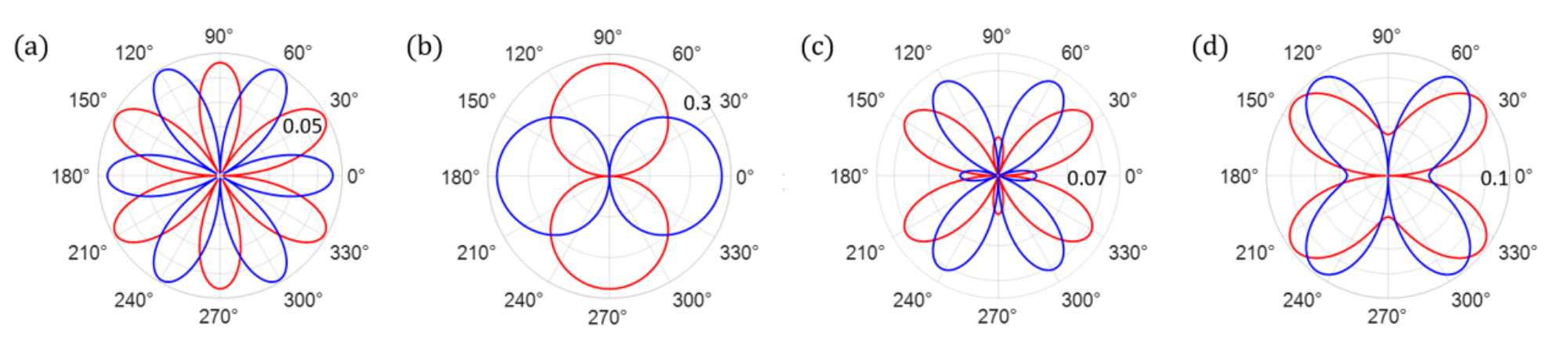}
\caption{Quasiparticle gap for $\Delta_{4x}$ (marked by solid blue lines) and $\Delta_{4y}$ (marked by solid red lines) pairings in the presence of linear and cubic RSOCs. Quasiparticle gap with only cubic RSOC of $\alpha_\text{CR}=0.1$ (a) and linear RSOC $\alpha_\text{LR}=0.4$ (b), respectively. (c), (d) Quasiparticle gap with coexisting linear and cubic RSOCs, where $\alpha_\text{CR}$ is fixed at 0.1, and $\alpha_\text{LR}$ takes values of 0.03, and 0.1, respectively. Common parameters,  $ \beta_0=0, \beta_1=-1, \mu=0, \varepsilon=0.35$; $ \Delta_0= 0.2$. } 
\label{SM:fig4}
\end{figure}

The competition between pairing channels can lead to other exotic phases. Pairing within the 2D representation (e.g.,~the $E_u$ pairing in Table~\ref{SM:stable1}) can lead to a nematic superconductor~\cite{SM-fu2014odd} characterized by broken rotational symmetry in the gap amplitude. Such a phase can be experimentally verified, for instance, recent studies have identified odd-parity nematic superconductivity in doped Bi$_2$Se$_3$~\cite{SM-matano2016spin,SM-yonezawa2017thermodynamic,SM-asaba2017rotational,SM-pan2016rotational,SM-du2017superconductivity,SM-tao2018direct}. 
Our calculations indicate that the $\Delta_{4}$ pairing can induce nematic superconductivity in this cubic Rashba spin-orbit coupled bilayer system. Figure~\ref{SM:fig4}(a) illustrates the anisotropic quasiparticle gap of $\Delta_{4x}$ and $\Delta_{4y}$ in the presence of pure cubic RSOC. $\Delta_4$ pairing spontaneously reduces the point group symmetry form $D_{4h}$ to $C_{2h}$. The quasiparticle gap displays six nodes, resulting in a nodal TSC. This six-fold nodal structure is a direct reflection of the complex, six-lobed anisotropy of the cubic RSOC spin texture away from the high-symmetry axes. For comparison, we also plot the quasiparticle gap in Fig.~\ref{SM:fig4}(b) when only linear RSOC is present. In this scenario, $\Delta_{4x}$ and $\Delta_{4y}$ have two nodes along the $k_y$ and $k_x$ axes, respectively. Figures~\ref{SM:fig4}(c) and \ref{SM:fig4}(d) demonstrate the quasiparticle gap with coexisting linear and cubic RSOCs. As the strength of the linear RSOC increases relative to a fixed cubic RSOC, the six nodes in the quasiparticle gap reduce to two nodes. 

\makeatletter
\let\SM@savedlabel\label
\let\SM@savedaddcontentsline\addcontentsline
\renewcommand{\addcontentsline}[3]{}
\renewcommand{\label}[1]{%
  \def\SM@tempa{#1}%
  \def\SM@tempb{LastBibItem}%
  \ifx\SM@tempa\SM@tempb
    \SM@savedlabel{SM:LastBibItem}%
  \else
    \SM@savedlabel{#1}%
  \fi
}
\makeatother
\makeatletter
\let\label\SM@savedlabel
\let\addcontentsline\SM@savedaddcontentsline
\makeatother
\end{document}